\documentclass[11pt, a4paper]{article}
\usepackage{jcappub}

\usepackage{color}
\usepackage{booktabs}
\usepackage{multirow}
\usepackage{amssymb}
\usepackage[export]{adjustbox}
\usepackage{floatrow}
\newfloatcommand{capbtabbox}{table}[][3.5in]
\newfloatcommand{capbfigbox}{figure}[][2.3in]
\usepackage{blindtext}
\usepackage{amsmath}
\usepackage{mathrsfs}
\usepackage{booktabs}
\usepackage{aas_macros}
\usepackage{epsfig}
\usepackage{graphicx}
\usepackage{subfigure}
\usepackage{comment}

\def\lsim{\mathrel{\raise.3ex\hbox{$<$\kern-.75em\lower1ex\hbox{$\sim$}}}}
\def\gsim{\mathrel{\raise.3ex\hbox{$>$\kern-.75em\lower1ex\hbox{$\sim$}}}}

\begin{document}

\title{A Multi-messenger Study of the Cygnus bubble}

\author[a,b]{Abhijit Roy,}
\author[c]{Giovanni Morlino,}
\author[d,h]{C Rishi,}
\author[d,e]{Jagdish C. Joshi,}
\author[e,f,g]{Soebur Razzaque,}
\author[d]{and Neelam Panwar}

\affiliation[a]{Gran Sasso Science Institute, via F. Crispi 7 -- 67100, L'Aquila, Italy}

\affiliation[b]{INFN/Laboratori Nazionali del Gran Sasso, via G. Acitelli 22, Assergi (AQ), Italy}

\affiliation[c]{INAF/Osservatorio Astrofisico di Arcetri, L.go E. Fermi 5, 50125 Firenze, Italy}

\affiliation[d]{Aryabhatta Research Institute of Observational Sciences (ARIES), Manora Peak, Nainital 263001, India}

\affiliation[e]{Centre for Astro-Particle Physics (CAPP) and Department of Physics, University of Johannesburg, PO Box 524, Auckland Park 2006, South Africa}

\affiliation[f]{Department of Physics, The George Washington University, Washington, DC 20052, USA}

\affiliation[g]{National Institute for Theoretical and Computational Sciences (NITheCS), Private Bag X1, Matieland, South Africa}
\affiliation[h]{Department of Applied Physics, Mahatma Jyotiba Phule Rohilkhand University, Bareilly, 243006, U.P., India}
\emailAdd{abhijit.roy@gssi.it (AR)}
\emailAdd{giovanni.morlino@inaf.it (GM)}
\emailAdd{rishictk@gmail.com (RC)}
\emailAdd{jagdish@aries.res.in (JCJ)}
\emailAdd{srazzaque@uj.ac.za (SR)}
\emailAdd{neelam\_1110@yahoo.co.in (NP)}

\newcommand\mc[1]{{\bf \color{magenta} #1}}
\newcommand\rs[1]{{\bf \color{cyan} #1}}

\abstract{
We perform a comprehensive investigation of particle acceleration at the termination shocks of massive star clusters, with a particular focus on the Cygnus region. Building on the particle-acceleration model for cluster-driven winds, we use recently available spectral and spatial measurements of ultra-high-energy $\gamma$-rays from the Cygnus bubble to place stringent constraints on the physical conditions in the acceleration zone. By jointly modelling the observed $\gamma$-ray spectrum and spatial morphology, we derive quantitative limits on key parameters such as the maximum particle energy, acceleration efficiency, magnetic-field strength, and shock geometry, and assess their consistency with a hadronic origin of the emission. These data allow us to refine the target gas profiles within and around the bubble, thereby improving estimates of the expected hadronic $\gamma$-ray and neutrino fluxes. Using this revised gas model, we compute the neutrino emission associated with the Cygnus bubble and evaluate the statistical significance of a potential neutrino signal from this region for current and near-future high-energy neutrino observatories. Taken together, these results provide a more precise characterisation of the particle-acceleration environment at cluster termination shocks and help clarify the role of young massive star clusters as potential Galactic PeVatrons and contributors to the observed diffuse high-energy neutrino and $\gamma$-ray emissions.}

\maketitle



\section{Introduction}

Supernova Remnants (SNRs) have long been regarded as prime candidates for the origin of Galactic Cosmic Rays (GCRs) \citep{Ahler2009PhRv_17A,Caprioli2012JC_38C,Schure2013MNR_4S,Bell13,Telez2013A_2T,Cardillo15,Brose2020A_9B,Sinit2023Univ_98S}. Recent observations by the Large High Altitude Air Shower Observatory (LHAASO) further support this scenario by indicating that sub-PeV protons can be accelerated in SNRs \citep{Cao25}. In addition, observations by LHAASO have revealed ultra-high-energy (UHE; $E_\gamma \ge 100$ TeV) gamma-ray emission from several Galactic sources, such as young massive star clusters (YMSCs), microquasars (MQs), and pulsar wind nebulae (PWNe), etc \citep{Cao2024ApJS_25C}. Understanding the acceleration of CR protons up to the knee feature at energies of order $\sim 3$ PeV remains a central problem, for which current UHE $\gamma$-ray observations provide crucial constraints \citep{Cao2023ARNPS41C}. 

Diffusive shock acceleration (DSA) is one of the leading mechanisms proposed for accelerating GCRs. In SNRs, DSA can operate at the shock fronts, which propagate into the surrounding interstellar medium (ISM). The resulting non-thermal particles may reach energies up to the PeV range under suitable physical conditions
\citep{Kry1977DoSS_6K,Bell1978MNRAS_47B,Bell1978MNRAS_43B,Drury1983RPP_73D,Blandford87,Malkov01}. Theoretical studies, that focused on resonant streaming instability \citep{Kulsrud69}, driven by the self-confinement of cosmic rays, concluded that the maximum proton energy achievable in a typical SNR is roughly an order of magnitude below the knee \citep{Lagage93a,Lagage93b}. By contrast, non-resonant streaming instability can excite small-scale magnetic perturbations in the upstream region, substantially amplifying the local magnetic field and thereby enabling the acceleration of GCRs to PeV-scale energies \citep{Bell04,Schure12}. Although type-Ia and core-collapse SNRs may transiently reach $\sim$PeV energies during their early evolutionary stages, their time-integrated effective maximum energy output is expected to contribute to the GCR spectrum predominantly up to only a few hundred TeV \citep{Cristofari20}. Furthermore, the observed $^{22}$Ne/$^{20}$Ne ratio in GCRs is enhanced by approximately a factor of five relative to the corresponding solar abundance \citep{Wiedenbeck81,Binns08,Prant2012A_80P}. This anomalous isotopic ratio cannot be satisfactorily explained under the assumption that GCRs are accelerated exclusively at SNR shocks. However, it can be accounted for if we hypothesize that a portion of these GCRs undergo acceleration within the stellar winds of Young Massive Star Clusters (YMSCs) \citep{gupta2020realistic, 2021MNRAS.508.1321T}.

Stellar winds from massive stars play a crucial role in accounting for the observed composition of GCRs \citep{Prant2012A_80P}. Moreover, such winds have long been proposed as potential sites for the acceleration of very high energy cosmic rays \citep{casse_1980ApJ_36C,cesa1983SSRv_73C, 1985ApJ...298..684W, volk1982cosmic, barretomota2026cosmicraydiffusionorigin}. Over the past decade, $\gamma$-ray observations in the GeV-PeV range from Galactic YMSCs -- such as Cygnus OB2 (Cyg OB2), BDS2003, NGC 3603, W40, Westerlund 1, and Westerlund 2, have been investigated to probe the presence of PeV protons within their stellar winds \citep{Yang2018A_77Y, Aharonian_2019NatAs561A,Saha2020ApJ_31S,Sun2020A_80S,Abe2021NatAs_465A,Albert2021ApJ_0A,Banik2022ApJ_30B,Blasi2023MNRAS_015B, 2022AdSpR..70.2685B}. In the present work, we focus on the multimessenger modelling of Cyg OB2 YMSC.

Cyg OB2 is among the most massive and compact OB associations known in the Milky Way galaxy \citep{kn2000A_..539K}, located at Galactic coordinates $l \sim 80.22039$, $b \sim 0.79661$. The association hosts several hundred massive stars distributed within a characteristic radius of $\sim 15$ pc \citep{kn2000A_..539K}. Within this stellar system, \cite{2015MNRAS.449..741W} identified 169 primary OB stars, including 52 O-type stars and 3 Wolf-Rayet (WR) stars. Based on stellar evolutionary models, they inferred that star formation in Cyg OB2 proceeded in a quasi-continuous manner over the past 1-7 Myr, with a peak activity epoch at approximately 4-5 Myr. The location of Cyg OB2, viewed nearly tangentially to the direction of Galactic rotation, complicates the separation of superposed structures along the line of sight (LOS) and thus hinders precise distance determination. An initial estimate placed the association at a distance of 2.1 kpc \citep{1966PROE....5..111R}. Subsequent photometric and spectroscopic observations of massive stars in Cyg OB2 yielded a revised distance of $1.7 \pm 0.2$ kpc \citep{torres1991MNRAS_T}. In particular, \cite{2019MNRAS.484.1838B}, in their study, report that the main Cyg OB2 group is likely farther away than previously assumed, implying that its stellar contents are intrinsically more luminous and more massive than earlier estimates suggested. More recent analyses, incorporating improved observational data and calibration techniques, have converged on a distance of approximately 1.6 -- 1.75 kpc \cite{2021MNRAS.508.2370Q, 2021MNRAS.502.6080O, haerer2025deciphering}.

Gamma-ray observations of Cyg OB2 with the Fermi Large Area Telescope (Fermi-LAT) indicate that the injection sites of CRs are spatially correlated with stellar winds and ionization fronts \citep{Ackermann2011_3A}. These CRs are detected within a region of approximately 50 pc, and their confinement time has been estimated to be $\sim 0.1$ Myr based on isotopic abundance measurements \citep{Ackermann2011_3A}. Within the regions of Fermi-LAT GeV emission, the High-Altitude Water Cherenkov (HAWC) observatory has also identified $\gamma$-ray emission in the 1-100 TeV range \citep{Abe2021NatAs_465A}. A TeV counterpart of the Cygnus Cocoon was additionally detected by the ARGO experiment \citep{Bartoli2014ApJ_52B}. More recently, the LHAASO has reported the detection of UHE $\gamma$-rays from the same region, but originating from a substantially larger spatial extent of at least 100 deg$^2$ \citep{lhaaso2024SciBu_49L}. In this field, photons with energies exceeding 400 TeV were observed, including two events with energies surpassing 1 PeV within $0.5^{\circ}$ of Cyg OB2 \citep{lhaaso2024SciBu_49L}. An earlier LHAASO result reported the detection of a 1.4 PeV photon, with the source LHAASO J2032+4102 potentially associated with Cyg OB2, PSR J2032+4127, or SNR G79.8+1.2 \citep{Cao2021Natur_33C}. 

\begin{figure}[!ht]
    \centering
    \includegraphics[width=0.65\linewidth]{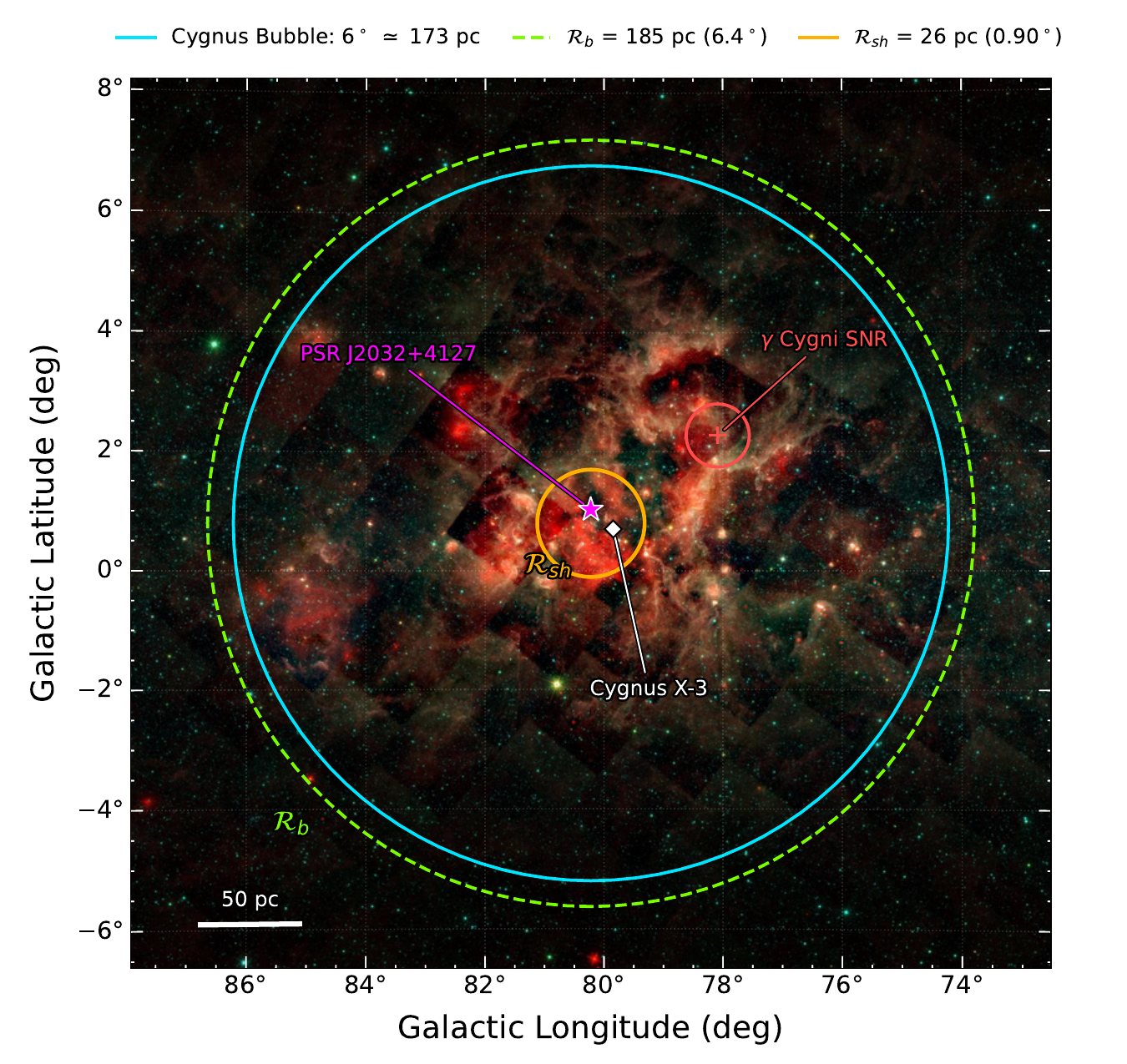}
    \caption{The Cygnus bubble region and the high-energy $\gamma$-ray sources observed in its vicinity are shown. The inner brown dotted circle indicates the TS region generated by the collective stellar winds, whereas the green dashed-dotted circle denotes the bubble region derived under the assumption of Bohm-like diffusion model parameters (see Tab. \ref{tab:ModelFit}). The blue circle represents the $6^{\circ}$ angular extension of the UHE $\gamma$-ray emission detected by LHAASO \citep{lhaaso2024SciBu_49L}. For the background image, we used an \textit{AllWISE} colour-composite dataset \cite{2010AJ....140.1868W, 2012wise.rept....1C}.}
    \label{fig:roi}
\end{figure}

In prior observations conducted in 2009, the Major Atmospheric Gamma Imaging Cherenkov (MAGIC) telescopes detected the TeV source J2032+4130 in this region with a significance of 5.6$\sigma$ \citep{Albert2008_5A}. Subsequent observations by the Very Energetic Radiation Imaging Telescope Array System (VERITAS) in 2014 confirmed this source with an increased significance of 8.7$\sigma$ \citep{Aliu2014ApJ_16A}. The origin of these $\gamma$-rays is likely linked to particle accelerators within the Cygnus bubble, such as the Cyg OB2 association, the pulsar PSR J2032+4127, the MQ Cygnus X-3, or the $\gamma$-Cygni SNR, as illustrated in Fig. \ref{fig:roi}. 

The presence of a rich population of OB stars in this system provides compelling evidence that star clusters can function as efficient accelerators of cosmic rays, with the resulting hadronic and leptonic interactions being probed through secondary $\gamma$-ray and neutrino emission \citep{butt2003astro_243B,Aharonian_2019NatAs561A}. However, the mere detection of secondary $\gamma$-ray emission from a YMSC does not, by itself, establish whether the underlying radiation mechanism is hadronic or leptonic, and thus does not unambiguously confirm the source as a cosmic-ray accelerator. High-energy $\gamma$-rays can be produced via leptonic processes such as Bremsstrahlung and Inverse Compton (IC) scattering, and for some YMSCs these leptonic channels appear to be the dominant emission mechanisms \citep{2023A&A...671A...4H}. By contrast, an analysis by \cite{neronov24}, based on a decade of IceCube data, reveals a $3\sigma$ excess in the neutrino flux from the Cygnus region, which can dominantly generated through hadronic interactions. This result provides additional support for the scenario in which the Cyg OB2 stellar cluster complex operates as a PeVatron, acting as a prominent accelerator of hadronic cosmic rays up to PeV energies.




In this work, we interpret the multi-messenger observations of the Cygnus bubble obtained by the Fermi-LAT, HAWC, ARGO, LHAASO, and IceCube detectors within the framework of the particle-acceleration model at the termination shock of YMSCs developed by \cite{Morlino21}. The origin of the $\gamma$-ray emission from this region has previously been investigated by several authors, including \cite{Blasi2023MNRAS_015B} and \cite{menchiari2024cygnus}, who applied a similar theoretical framework to the comparatively smaller Cygnus cocoon, constrained at the time by the limited availability of observational data. In addition, recently \cite{haerer2025deciphering} modelled the extended $\gamma$-ray emission of the Cygnus bubble by invoking a lepto-hadronic emission scenario and positing the presence of an evolved SNR with an estimated age of $\sim 50$ kyr in the Cygnus region. In this same paper, the author questions the existence of a large-scale termination shock due to the wide distances between stars in Cyg OB2. Recent simulations of compact stellar clusters \cite{2026A&A...709A..11V, 2024MNRAS.532.2174V} suggest that the formation of a quasi-spherical, large-scale wind termination shock is non-trivial. Achieving such a morphology requires a nearly homogeneous distribution of mechanical power among dozens of massive stars confined within a very compact volume ($\leq1$ pc). However, in our considered framework, even if a single stellar cluster is unable to generate a coherent, large-scale termination shock, termination shocks associated with individual stellar winds, or the superposition of shocks arising from several distinct star-forming regions, may nonetheless provide efficient sites for particle acceleration to very high energies \cite{2000APh....13..161K, 2004A&A...424..747P}. In contrast, in the present study, we exploit recently available spectral and spatial measurements of UHE $\gamma$-rays from the Cygnus bubble to place stringent constraints on the model parameters and to derive quantitative information on the underlying particle-acceleration sites. We predict a neutrino flux based on the same model that fits the $\gamma$-ray data and estimate the statistical significance of neutrino detection by the IceCube and KM3NeT neutrino telescopes.


The structure of the paper is as follows. In Section \ref{sec:CRdist}, we provide a concise description of the adopted theoretical model, followed by its implications for multi-messenger emission in Section \ref{sec:mmEmission}. In Section \ref{sec:result}, we outline our fitting procedure and compare the model predictions with the available observational data. In Section \ref{sec:EventRate}, we present the expected neutrino event rates from the source for the IceCube and KM3NeT neutrino telescopes, considering different exposure times. Finally, we summarise and discuss our conclusions in Section \ref{sec:conclusion}.

\section{Particle Acceleration in Star Clusters}
\label{sec:CRdist}
In recent years, YMSCs have gained more attention as a potential source of CRs in our Galaxy because of the detection of VHE to UHE $\gamma$-rays from these regions. Nevertheless, the precise location where particle acceleration occurs remains poorly understood. It is hypothesised that the shock wave generated at the core of YMSCs, due to the interaction between stellar winds and/or the magnetic turbulence resulting from such collisions, could serve as the origin of this acceleration. An alternative hypothesis, recently investigated by \citep{Morlino21, Blasi2023MNRAS_015B}, suggests that the acceleration may occur at the Termination Shock (TS) region. This region is formed by the interaction between the collective winds from young, compact stellar clusters and the ISM. The structure of a wind-driven bubble produced by compact, young, and powerful star clusters has been extensively characterised in previous studies \citep{menchiari2024cygnus, Morlino21, Blasi2023MNRAS_015B}. Therefore, we do not elaborate on the structural details here, as a thorough description can be found in the above references. However, for the sake of completeness, we aim to outline several important parameters that characterise these bubbles.

\subsection{The Acceleration Model}
\label{sec:bubble}
Let us first consider that the core of the YMSC is very compact, with all the stars concentrated within a radius \( \mathcal{R}_c \). The collective stellar wind of the YMSC, which has a velocity \( u_{\rm sw} \), excavates a bubble-like structure having a density profile of,
\begin{equation}
    \varrho(r) = \frac{\dot{M}_{\star}}{4\pi r^2 u_{\rm sw}},~~~ r > \mathcal{R}_c
\end{equation}
where $\dot{M}_{\star}$ is the mass loss rate from the collective wind, which is related to the total luminosity of the system as $\mathcal{L}_{\rm sw} = \dot{M}_{\star}u_{\rm sw}^2/2$. The interaction of stellar winds is bounded by the TS at $\mathcal{R}_{sh}$, while the supersonic collective wind creates a forward shock at $\mathcal{R}_b$ from its interaction with the ISM that has a constant density of $\varrho_0$. The location of the forward shock, specifically the radius of the bubble and the TS region, is well approximated in \cite{Morlino21, menchiari2024cygnus}, and we are using the same approximation here. The formulae for \( \mathcal{R}_b \) and \( \mathcal{R}_{\rm sh} \) are as follows:

\begin{equation}
    \mathcal{R}_b = \left(\frac{125}{154\pi}\frac{\mathcal{L}_w}{\varrho_0}\right)^{1/5} \mathcal{T}_{age}^{3/5}
\end{equation}
and 
\begin{equation}
    \mathcal{R}_{sh} = \left(\frac{(3850\pi)^{2/5}\dot{M}_{\star}u_{sw}}{28\pi}\right)^{1/2} \mathcal{L}_{sw}^{-1/5} \varrho_0^{-3/10} \mathcal{T}_{age}^{2/5}
\end{equation}
where \( \mathcal{T}_{age} \) is the age of the YMSC and the other parameters have their usual meaning. The mechanism of radiative cooling is also not taken into account in this analysis, as it is deemed negligible within the timescale of a few Myr of \( \mathcal{T}_{age} \), during which the system is considered to undergo complete adiabatic expansion \citep{Morlino21}. Although radiative cooling and the presence of dense clumps may have some effect on the size of the bubble and TS region \citep{menchiari2024cygnus, blasi2024different, Blasi2023MNRAS_015B}, they do not alter the overall picture of cosmic ray transport in the YMSC, and the assumption of adiabatic expansion remains valid. The impact of energy loss has been studied by \cite{Blasi2023MNRAS_015B}; they conclude that it does not significantly modify the overall spectral shape. However, it can influence the spatial distribution of low-energy ($\lesssim 50$ TeV) particles if the gas density of the wind is higher than $10\;\rm cm^{-3}$. Recently, \cite{blasi2024different} also shows that the energy loss for heavy nuclei due to spallation reactions is significant even at lower densities; hence, YMSC does not release into the ISM a significant amount of heavy elements, with the possible exception of the late phases when the circumstellar gas is swept up.

In this modelling framework, we focus primarily on the acceleration of hadronic particles, specifically protons, and neglect leptonic contributions from electrons. While a minor electron contribution cannot be ruled out, it is unlikely to dominate the emission above energies of a few TeV. Indeed, based on Swift-XRT observations, \cite{2023ApJ...950..116G} showed that, assuming an equipartition magnetic field of $\sim 20\;\mu\mathrm{G}$, the upper limit on the leptonic IC component does not exceed approximately one quarter of the $\gamma$-ray flux at 1 TeV. Similarly, analyses of Chandra X-ray observations \cite{colombo2018diffusexrayemissioncygnus} indicate that leptonic mechanisms, such as synchrotron emission and inverse Compton scattering, make a comparatively minor contribution to the large-scale diffuse emission. Although DSA is generally efficient at accelerating hadronic particles, the $>3\sigma$ detection of neutrinos from the Cygnus cocoon by \cite{neronov24} further supports a predominantly hadronic origin of the high-energy emission. Nevertheless, we explicitly evaluate the possible leptonic contribution. The corresponding calculations are presented in Appendix \ref{seca:Lepton}, where we show that the leptonic component remains subdominant unless DSA work in a significantly different way from SNR shocks (see Appendix \ref{seca:Lepton}).

The energy-dependent radial distribution of non-thermal protons per unit energy and per unit volume, $N(r, E) = 4\pi p^2 N(r, p) dp/dE$ accelerated at the TS of the bubble via DSA can be found by solving the 1D transport equation in spherical geometry as described in \cite{Morlino21}.

\begin{equation}
\frac{1}{r^{2}} \frac{\partial}{\partial r}
\left[ r^{2} \mathcal{K}(r,p)\, \frac{\partial N}{\partial r} \right]
- u(r)\, \frac{\partial N}{\partial r}
+ \frac{1}{r^{2}} \frac{d}{dr}\!\left[ r^{2} u(r) \right]
\frac{p}{3} \frac{\partial N}{\partial p}
+ \mathcal{S}(r,p) = 0 ,
\label{eq:transport}
\end{equation}
where $\mathcal{K}(r, p)$ is the diffusion coefficient, $u(r)$ is the wind speed, $\mathcal{S}(r,p)$ represents the particles injection term, considered to only take place at the TS ($r = \mathcal{R}_{sh}$) according to the distribution:

\begin{equation}
    \mathcal{S}(r,p) = \frac{\zeta_{inj}\;\varrho_{sw}\;u_{sw}}{4\pi p^2_{inj}} \delta(p - p_{inj}) \delta(r-\mathcal{R}_{sh})
\end{equation}
where $\zeta_{inj}$ denotes the fraction of particle flux engaged in the acceleration mechanism at $p = p_{inj}$, $\varrho_{sw}$ signifies the density of the cold plasma located immediately upstream of the termination shock, and $u_{sw}$ represents the velocity of the cold plasma.


The transport equation can be solved numerically or analytically, as shown in \cite{Blasi2023MNRAS_015B, blasi2024different} and \cite{Morlino21}. Fitting the observed data to the model output often requires running the model with many different sets of input parameters. Therefore, here we are utilizing a parametric fitted distribution derived by \cite{menchiari2024cygnus} from the full analytic solution of the transport equation as detailed in \cite{Morlino21}. The solution is derived independently for three distinct regions: within the cold wind ($N_{\rm sw}$), inside the hot bubble ($N_{\rm b}$), and outside the bubble ($N_{\rm out}$) in the ISM. The final solution is then integrated by considering the conservation of flux at the boundary of the TS and the bubble, with the boundary condition assuming that the flux is zero at the cluster centre as implied by spherical symmetry and aligns with the diffuse GCR flux ($N_{\rm gal}$) at a far distance. The spatial distribution of particle flux, accelerated in such a system, is thus reduced to the following form:


\begin{subequations}
\begin{flalign}
\label{eq:Nrp_a}
&N_{\rm sw}(r,p) \simeq N_{\mathrm{sh}}(p)\,\exp\!\left[-\frac{u_{sw}(\mathcal{R}_{sh} - r)}{\mathcal{K}_{sw}(p)}\right]
\\[10pt]
\label{eq:Nrp_b}
&N_{b}(r,p) =
\frac{
N_{\mathrm{sh}}(p)\!\left[e^{\alpha} + \beta\!\left(e^{\alpha_{b}} - e^{\alpha}\right)\right]
+ \xi_H N_{\mathrm{gal}}(p)\,\beta\!\left(e^{\alpha} - 1\right)
}{
1 + \beta\!\left(e^{\alpha_{b}} - 1\right)
}
\\[10pt]
\label{eq:Nrp_c}
&N_{\mathrm{out}}(r,p) =
N_{b}(\mathcal{R}_{b},p)\,\frac{\mathcal{R}_{b}}{r}
+ \xi_{H} N_{\mathrm{gal}}(p)\!\left(1 - \frac{\mathcal{R}_{\mathrm{sh}}}{r}\right)
\end{flalign}
\end{subequations}
where $\xi_{H}= 1.288$ denotes the Nuclear Enhancement Factor (NEF), which accounts for the contribution of heavy nuclei contained in the GCR population \citep{2009APh....31..341M}. In the present treatment, we neglect any additional contribution from heavy elements that may be accelerated in YMSCs \citep{blasi2024different}. The NEF related to the heavy elements present both in the ISM and in the YMSC environment also enhances the multi-messenger signal and is accounted for later in Equ. \ref{equ:mmE}. The exponents $\alpha$ and $\beta$ are two variables that take the following form:
\begin{subequations}
\begin{flalign}
\label{eq:rp_a}
&\alpha(r,p) = \frac{u_b \mathcal{R}_{sh}}{\mathcal{K}_b(p)}\left(1-\frac{\mathcal{R}_{sh}}{r}\right)
\\[10pt]
\label{eq:rp_b}
&\beta(p) =\frac{\mathcal{K}_{out}(p) \mathcal{R}_b}{u_b \mathcal{R}_{sh}^2}
\end{flalign}
\end{subequations}
and $\alpha_b$ is the value of $\alpha(r,p)$ at bubble radius i.e., at $r = \mathcal{R}_b$. The wind velocity at the downstream bubble, denoted by $u_b$, is compressed by a factor of 4 (assuming a strong shock) relative to the upstream wind speed $u_{\rm sw}$ \citep{blasi2024different}. This compression ratio applies to a strong shock with a high Mach number; the condition readily available in the Cygnus bubble. Three different spatially constant diffusion coefficients are considered here: $\mathcal{K}_w(p)$ for the wind, $\mathcal{K}_b(p)$ for the bubble, and $\mathcal{K}_{\rm out}(p)$ for the ISM. While the final expression for the particle distribution at TS is reduced to:


\begin{equation}
    N_{\rm sh}(p) = \frac{3 \varrho_{sw} u_{\rm sw}^2 \epsilon_{\rm cr}}{4\pi \Lambda_p (m_pc)^3 c^2} \left(\frac{p}{m_pc}\right)^{-s} e^{-\Gamma(p)}
\end{equation}
where the percentage of wind luminosity ($\mathcal{L}_{\rm sw}$) channelled into the accelerated particles is denoted by $\epsilon_{\rm cr}$, and $\Lambda_p = \int_{x_{\rm inj}}^{\infty} x^{2-s}e^{-\Gamma(x)} (\sqrt{1+x^2} -1)\;dx \simeq {\rm ln}(p_{\rm max}/m_pc) \sim 13$, with $x=p/m_pc$ \citep{Morlino21}. The exponential part of the shock distribution function can be fitted with an analytical form and contains information about particle diffusion in both the upstream and downstream of the shock.

\begin{equation}
    e^{-\Gamma(p)} \simeq \left[1 + a_1\left(\frac{p}{p_{\rm max}}\right)^{a_1}\right] {\rm exp}\left[-a_3\left(\frac{p}{p_{\rm max}}\right)^{a_4}\right]
    \label{eq:shDist}
\end{equation}
The $p_{\rm max}$ in Equ. \ref{eq:shDist} is not the true maximum momentum that can be accelerated in YMSCs; rather, it is the maximum momentum, determined by upstream conditions only, and the fit parameters are listed in Tab. \ref{tab:diffusion_parameters}.

\begin{table}
\centering
\renewcommand{\arraystretch}{1.25}
\setlength{\tabcolsep}{5pt}
\begin{tabular}{lcccc}
\hline\hline
Model & $a_{1}$ & $a_{2}$ & $a_{3}$ & $a_{4}$ \\
\hline
Kolmogorov & 10.00 & 0.308653 & 22.0241 & 0.43112 \\
Kraichnan   & 5.00  & 0.448549 & 12.5200 & 0.642666 \\
Bohm        & 8.94  & 1.29597  & 5.31019 & 1.13245 \\
\hline
\end{tabular}
\caption{The parameter values that represent the full solution under different diffusion conditions adopted from \cite{menchiari2024cygnus}.}
\label{tab:diffusion_parameters}
\end{table}

\begin{figure}
    \centering
    \includegraphics[width=0.65\linewidth]{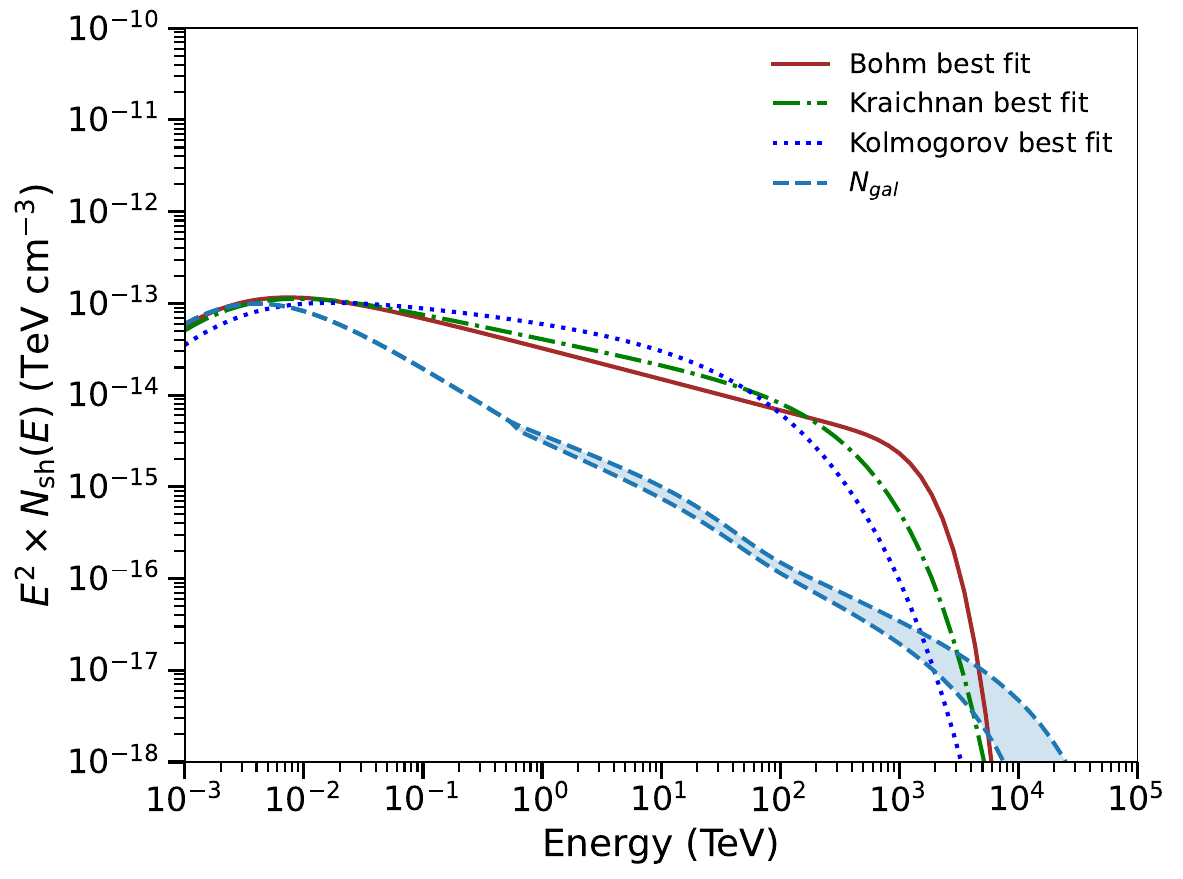}
    \caption{The distribution of proton flux in the TS region, accelerated by Cyg OB2, and the GCR proton flux of the local interstellar spectrum is depicted here. The proton flux at the TS is derived using the best-fit parameters set from different diffusion in the fitting procedure (see Tab. \ref{tab:ModelFit}). For comparison, the shaded region in the figure represents the uncertainty in the high-energy GCR measurements \citep{2023PhRvL.131o1001C}.}
    \label{fig:FcrVsE}
\end{figure}

In YMSCs, stars exhibit varying rates of mass loss and wind velocities. When these stellar winds interact, a fraction of their combined kinetic energy is assumed to be converted into thermal and magnetic energy. However, the precise location of this conversion process remains uncertain. Following the methodology of \cite{Morlino21}, we assume that a fraction, denoted as $\eta_B$, of the wind's kinetic energy is transformed into magnetic energy in the upstream wind. Hence, the magnetic field at the upstream wind is
\begin{equation}
    \delta \mathcal{B}_{sw} = \sqrt{\eta_B \frac{\dot{M}_{\star} u_{sw}}{2r^2}}
\end{equation}
While at the downstream of TS, the magnetic field is assumed to be compressed by a factor $\sqrt{11}$ for a strong shock. The diffusion of particles strongly depends on magnetic turbulence, which is not well understood. In this calculation, we will consider all known types of diffusion, including Bohm, Kolmogorov, and Kraichnan, which have the following forms:

\begin{subequations}
\begin{flalign}
    &\mathcal{K}_{\rm Bohm}(E) = \frac{v_p \ell_B}{3}
    \label{eq:kbohm}
\\[10pt]
    &\mathcal{K}_{\rm Kol}(E) = \frac{v_p \ell_B^{1/3} L_{\mathrm{inj}}^{2/3}}{3}
    \label{eq:kkol}
\\[10pt]
    &\mathcal{K}_{\rm Kra}(E) = \frac{v_p \ell_B^{1/2} L_{\mathrm{inj}}^{1/2}}{3}
    \label{eq:kkra}
\end{flalign}
\end{subequations}
Here, the particle velocity is defined as $v_p$, the Larmor radius is given by $\ell_B = cp/(e\delta \mathcal{B}_{\rm inj})$, and the turbulence injection scale, $L_{\mathrm{inj}}$, is assumed to be of the same order as the YMSC core (1--2 pc).

The upper limit of energy within stellar clusters is influenced by the particular diffusion mechanism under consideration. The maximum energy values for the three diffusion types analysed in this study can be articulated in accordance with the equations delineated in \cite{menchiari2024cygnus}.

\begin{equation}
    E_{\rm max}^{\rm Bohm} \simeq 10 \sqrt{\frac{\eta_B}{0.1}} \left(\frac{\dot{M}_{\star}}{10^{-4}M_{\odot}yr^{-1}}\right)^{-1/4} \left(\frac{\mathcal{L}_w}{10^{39} \mathrm{erg\;s}^{-1}}\right)^{3/4} \, \mathrm{PeV}
    \label{eq:maxEbohm}
\end{equation}

\begin{equation}
\begin{split}
E_{{\mathrm{max}}}^{\mathrm{Kol}} \simeq\; & 1.2 
\sqrt{\frac{\eta_B}{0.1}}
\left(\frac{\dot{M}_{\star}}{10^{-4}\,M_{\odot}\,\mathrm{yr}^{-1}}\right)^{-3/4}
\left(\frac{\mathcal{L}_{sw}}{10^{39}\,\mathrm{erg\,s}^{-1}}\right)^{37/20} \\
& \times
\left(\frac{\varrho_0}{20\,m_p\,\mathrm{cm}^{-3}}\right)^{-3/5}
\left(\frac{\mathcal{T}_{\mathrm{age}}}{3\,\mathrm{Myr}}\right)^{4/5}
\left(\frac{L_c}{2\,\mathrm{pc}}\right)^{-2}
\,\mathrm{PeV}
\end{split}
\label{eq:maxEkol}
\end{equation}

\begin{equation}
\begin{split}
E_{{\mathrm{max}}}^{\mathrm{Kra}} \simeq\; & 2.84
\sqrt{\frac{\eta_B}{0.1}}
\left(\frac{\dot{M}_{\star}}{10^{-4}\,M_{\odot}\,\mathrm{yr}^{-1}}\right)^{-1/2}
\left(\frac{\mathcal{L}_{sw}}{10^{39}\,\mathrm{erg\,s}^{-1}}\right)^{13/10} \\
& \times
\left(\frac{\varrho_0}{20\,m_p\,\mathrm{cm}^{-3}}\right)^{-3/10}
\left(\frac{\mathcal{T}_{\mathrm{age}}}{3\,\mathrm{Myr}}\right)^{2/5}
\left(\frac{L_c}{2\,\mathrm{pc}}\right)^{-1}
\,\mathrm{PeV}
\end{split}
\label{eq:maxEkra}
\end{equation}

The spectral distribution of the particle population at the shock front, corresponding to the fitted parameters of the three adopted diffusion models (see Tab. \ref{tab:ModelFit}), is presented in Fig. \ref{fig:FcrVsE}. The description of the fitting procedure is described in detail in Sec. \ref{sec:result}. The local interstellar spectrum of diffuse GCR protons is also shown in the figure for comparison. Here we used the superimposition of two distribution functions, one for lower energy with $E \le 550$ GeV \citep{2015ApJ...815..119V} and another for higher energy $E > 550$ GeV \citep{2023PhRvL.131o1001C}. The shaded region in the figure represents the uncertainty in the high-energy GCR measurements.

\section{Multi-Messenger Emissions in Cygnus Bubble}
\label{sec:mmEmission}
The multi-messenger signals from YMSCs will reveal the nature of particle acceleration occurring there. These secondary particles, $\gamma$-rays and neutrinos ($\nu$), are produced in the YMSCs due to inelastic interactions between the accelerated particles and the target gas density present there, specifically through $p-p$ interactions. The differential flux of particles on Earth, emitted from a volume \( V \) and located at a distance \( d \) ($\sim 1.65\;\rm kpc$ assumed in this study) can be calculated using

\begin{equation}
    F_{\gamma, \nu}(E_{\gamma, \nu}) = \frac{c\xi_{n}}{4\pi d^2}\int\int N_{\rm cr}(r, E_p) n(r)\frac{d\sigma(E_p, E_{\gamma, \nu})}{dE_p} dE_p dV
    \label{equ:mmE}
\end{equation}
where $N_{\rm cr}(r, E_p)$ is the radially dependent accelerated particle flux, $n(r)$ is the spatial gas density profile, and $d\sigma(E_p, E_{\gamma, \nu})/dE_p$ is the differential cross section of $\gamma$-ray and neutrino production from $p-p$ interaction. The parameterisation of \cite{Kafexhiu14} and \cite{Koldobskiy21} is used here for the differential cross section to calculate the resulting emission of $\gamma$-ray and neutrino from the $p-p$ interaction. The enhancement of the multi-messenger contribution arising from heavy elements present both in the ISM and in the YMSC environment is quantified by the parameter \(\xi_{n}\) (NEF = 1.428). \citep{2009APh....31..341M}. The value of the NEF adopted here is for 10 GeV/nucleon; however, at higher energies, the factor can increase slightly \citep{2009APh....31..341M, Kachelriess14, Roy23}.
For the GCR component, we also adopt a constant NEF of 1.84, corresponding to an incident primary particle energy of 10 GeV/nucleon. This factor can rise to approximately 2 at a primary particle kinetic energy of 1000 GeV/nucleon \citep{2009APh....31..341M}, implying a relative increase of about 8.69\%. Nevertheless, this resulting enhancement in the hadronic $\gamma$-ray yield can be partially compensated by $\gamma-\gamma$ absorption, which becomes increasingly important at such energies, reaching its maximum around 2 PeV and leading to an attenuation of the flux by $\sim 19\%$.


\subsection{Gas Density Distribution}
The estimation of gas density is very crucial in this calculation. Hence, we derived the molecular gas density from CO line emission, the atomic gas density from 21-cm $H_{\rm I}$ measurements, and the ionised gas density from radio continuum data. These multi-phase tracers allowed us to estimate the density of the region and evaluate its spatial variations. We adopted a distance of $d_0=1.65$~kpc centered around $(l_0,b_0)=(80.22^\circ,0.7966^\circ)$ for Cygnus OB2, with a cocoon radius of $R_{\rm b}=150$~pc, for calculation of density.

For the molecular gas probe, we obtained the $^{12}$CO (J=1-0), $^{13}$CO (J=1-0), and C$^{18}$O (J=1-0) calibrated data from the 13.7-m millimetre-wave radio telescope, Purple Mountain Observatory (PMO), located in Delingha.
The noise level $rms$, which is about 0.5 K for $^{12}$CO (J=1-0) and 0.3 K for $^{13}$CO (J=1-0) and C$^{18}$O (J=1-0), is uniform across the entire covered area. All data cubes were of size $3.5^{\circ}$ centered on the Cyg OB2 Association ($l\ \sim \ 80.22039,\ b\ \sim \  +0.79661$). 
The 21 cm $H_{\rm I}$ line, which traces atomic gas, was taken from the Canadian Galactic Plane Survey (CGPS; \cite{2003AJ....125.3145T}) with area coverage similar to that of the molecular data. The ionised gas distribution associated with the Cygnus OB2 bubble was estimated using the free-free emission measure EM(l, b) extracted from component separation of Planck, WMAP, and 408 MHz \citep{2016A&A...594A..10P}.

To calculate the 3D density of molecular hydrogen in Cygnus, we follow the method outlined by \cite{2008ApJ...679..481P}, using the CO molecular lines. We separated the clouds in distance according to the analysis carried out by \cite{haerer2025deciphering}. We calculated the gas volume density in the region while explicitly excluding foreground and background cloud contributions. To find the number density of molecular hydrogen (n$_{\rm H_{2}}$), we divide the column density of individual clouds by their sizes taken from \citep{2024AJ....167..220Z}. We identify several clouds in the vicinity of the bubble with properties comparable to those reported by \cite{haerer2025deciphering}; accordingly, we adopt the same cloud parameters. However, given the 5\% uncertainty in the distance measurements, the clouds’ line-of-sight positions relative to the bubble centre are not tightly constrained. Consequently, these clouds may lie at any distance inside the bubble, which is not precisely known, and we therefore treat their distances as free parameters in the modelling.

For neutral gas, we construct a 3D neutral gas model based on the method used by \cite{haerer2025deciphering}, where they associate the neutral gas with
known molecular cloud complexes: the Cygnus Rift at a distance of about 1 kpc and the Cygnus SFR at about 1.7 kpc. They calculated the 3D column density model, which is converted into volume number density by assuming a Gaussian distribution of gas along the line of sight. The column density at each position was divided by the effective line-of-sight depth, determined by the Gaussian dispersion, to obtain the corresponding volume number density ($n_{\rm H_I}$). The density at each distance was then weighted by the Gaussian profile, such that it peaks at the adopted central distance and decreases with increasing distance from the centre.

For ionised emission, the EM map from the Planck survey is used \citep{2016A&A...594A..10P}. Assuming uniform electron density along each line of sight, the electron density was estimated as
\begin{equation}
n_{\rm H_{II}}=\sqrt{\frac{{\rm EM}(l,b)}{L(l,b)}},
\end{equation}
where $L(l,b)$ is the geometrical path length through the spherical bubble, given by
\begin{equation}
L(l,b)=2\sqrt{R_{\rm b}^2-r_{\rm proj}^2(l,b)},
\end{equation}
with
\begin{equation}
r_{\rm proj}^2(l,b)=d_0^2\left[
\sin^2(l-l_0)+\sin^2(b-b_0)
\right].
\end{equation}
The resulting electron density map was subsequently extended into three dimensions by assigning the corresponding density to voxels satisfying $r_{\rm 3D}\leq R_{\rm b}$, where
\begin{equation}
r_{\rm 3D}^2 =
d^2\cos^2 b\,(l-l_0)^2
+d^2(b-b_0)^2
+(d-d_0)^2.
\end{equation}
The 3D distribution was sampled from 500 to 2500~pc in steps of 10~pc, with zero density assigned outside the adopted spherical volume. The analysis indicates a substantial contribution of ionised gas despite a very low atomic gas density.

So, the total 3D average number density $n_T$ within the bubble, excluding the discrete cloud contribution, is
\begin{equation}
    n_T = n_{H_I} + 1.17n_{H_{II}}
\end{equation}
where, $\mathrm{n_{H_2}}$ is the molecular hydrogen density, $\mathrm{n_{HI}}$ is the atomic hydrogen density and $\mathrm{n_{HII}}$ is the ionized gas density. Factor 1.17 is considered for the case when both H and He are fully ionised \cite{2020MNRAS.498.6083E}. The line-of-sight integrated spatial density distribution of the three species is plotted in Fig. \ref{fig:2Dgas}.

\begin{figure}
    \centering
    \includegraphics[width=0.65\linewidth]{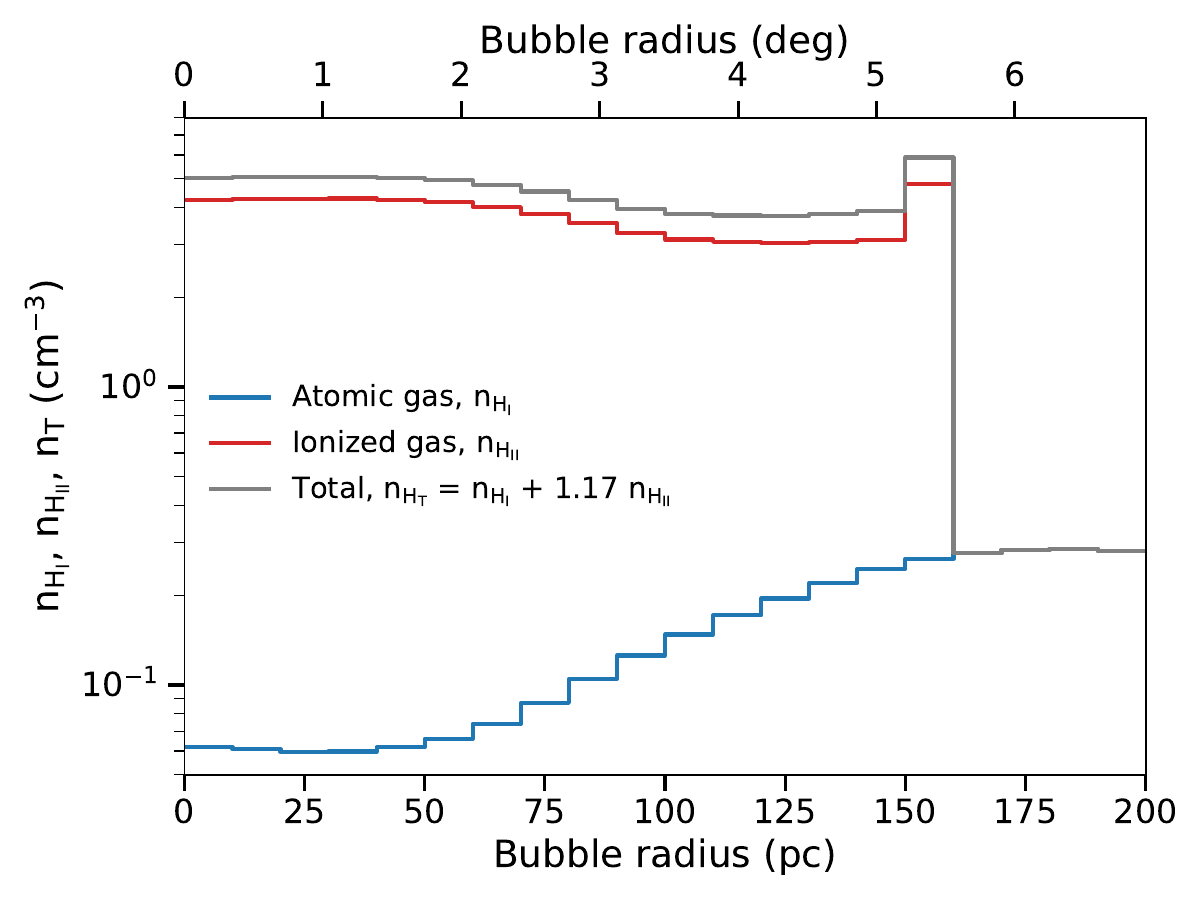}
    \caption{The density distribution of atomic (H$_I$) and ionised (H$_{II}$) hydrogen, as well as their combined contribution, within the Cygnus region, is shown here. Owing to the compact and spatially inhomogeneous distribution of molecular clouds within the bubble, their contribution is not included in this plot.}
    \label{fig:density}
\end{figure}

\subsection{Surface Brightness}
The LHAASO collaboration has presented the Surface Brightness (SB) profile of the Cygnus bubble in three energy intervals (2-20, 25-100, and $> 100$ TeV). For comparison, we have also computed the corresponding SB profile in the same energy bands based on our modelling, employing the formula given below.

\begin{equation}
    \mathrm{SB}_{\gamma}(l) = \frac{4\pi d^2}{\Omega_{\mathrm{annuli}}} \int E_{\gamma} F_{\gamma}(E_{\gamma}, l) dE_{\gamma}
\end{equation}
where $\Omega_{\mathrm{annuli}}$ is the angular area of the annulus area of the LOS observation, and $F_{\gamma}(E_{\gamma}, l)$ is the corresponding differential flux projected along the LOS.

The flux in the projected two-dimensional LOS radii is calculated following the methodology outlined in \cite{Rensburg18}. The radiation within a given annulus can be determined if the volume of the intersection between a specified hollow cylinder and the relevant spherical shells is known. The total radiation associated with a particular LOS, or annulus, is obtained by summing the radiation density over all contributing segments and subsequently multiplying this sum by the volume of intersection between the corresponding hollow cylinder and the spheres. 


\subsection{Neutrino Event Rates and Atmospheric Background}
\label{subsec:nurate}
Unlike $\gamma$-rays, neutrinos are produced through hadronic interactions only. Therefore, detecting neutrinos from a source can confirm it as a source of hadronic GCRs. In this context, we have also calculated the neutrino event rate at Earth from the Cygnus region, as expected by neutrino observatories such as IceCube and KM3NeT in the energy range from 1 to 1000 TeV. The neutrino event accumulated over a time period \( T \) from a source located at a declination \( \delta ec \) can be expressed in terms of the following equation. 

\begin{equation}
\mathcal{N}_{\rm event} = \int_{0}^{T} dt \int \frac{dF_{\nu}}{dE_{\nu}} A(E_{\nu}, \delta ec) dE_{\nu} 
\end{equation}
where ${dF_{\nu}}/{dE_{\nu}}$ is the differential neutrino flux and $A(E,\delta ec)$ is the effective area of the detector at the declination angle of the source. The effective areas of IceCube and KM3NeT are different. For the IceCube, we are using the effective area from \cite{2014ApJ...796..109A} for a source having $\delta ec   = 40.09^\circ$. Whereas the effective area for KM3NeT is taken from \cite{km3netAeff}, which is characterised by the source visibility at a given $\delta ec$ from \cite{km3netvis}.


The atmospheric neutrinos are the major background in the detection of astrophysical neutrinos. The atmospheric neutrino flux above 1 TeV can be approximated well by a power law equation. \cite{2024ApJ...969..161G} finds the normalisation and the index of the distribution by fitting the zenith-averaged differential atmospheric muon neutrino flux from 100 GeV to 400 TeV from \cite{2011PhRvD..84h2001A} and extrapolated linearly following a power-law dependence up to 1 PeV, which can be represented in the following form:
\begin{equation}
\frac{dN_{\nu_{\mu}}^{atm}}{dE_{\nu}} = 2.64\times10^{-8} E_{\nu}^{-3.4}\;\;\mathrm{cm^{-2} s^{-1} sr^{-1} TeV^{-1}}
\end{equation}

The background events due to atmospheric neutrinos can't be segregated on an event-by-event basis; they can only be subtracted on a statistical basis. The expected atmospheric neutrino events for the IceCube and KM3NeT around the search cone of the source under investigation are calculated as:
\begin{equation}
\mathcal{N}_{bkg} = 2 \int_{\Delta\Omega}\int_{0}^{T} dt \int \frac{dN^{atm}_{\nu}}{dE_{\nu}} A(E_{\nu}, \delta) d\Omega dE_{\nu} 
\end{equation}
The angular resolution of the detector limits the aperture of the search cone ($\Delta\Omega$), limited by the Point Spread Function (PSF) of each detector. An opening angle of $\pm 6^{\circ}$ is considered here, which is rather an optimistic value.

\section{Results and discussions}
\label{sec:result}
 We employed a Bayesian fitting \textsc{Python} module \textsc{Nautilus} \citep{nautilus} to determine the optimal combination of input parameters $\mathcal{P}$ that best reproduces both the observed Spectral Energy Distribution (SED) and the SB profile in three energy bands of Cygnus bubble, $\mathcal{D}$. The best-fit is achieved by maximising the likelihood function $\mathscr{L}_i(\mathcal{D}_i \mid \mathcal{P})$ for each set of input parameters $\mathcal{P}$, given the corresponding model predictions $\mathcal{M}$.
 
\begin{subequations}
\begin{align}
\mathscr{L} &= \prod_{i=1}^{4} \mathscr{L}_i(\mathcal{D}_i \mid \mathcal{P}) \\
\ln \mathscr{L} &= \sum_{i=1}^{4} \ln \mathscr{L}_i(\mathcal{D}_i \mid \mathcal{P})
\end{align}
\end{subequations}

The goodness-of-fit is, however, not straightforward to infer directly from the value of $\mathscr{L}$ or $\ln \mathscr{L}$, although a correspondence with the associated $\chi^2$ statistic can be established. It should be stressed that the value of the maximized likelihood function, $\ln \mathscr{L}_i(\mathcal{D}_i \mid \mathcal{P})$, for a given set of input parameters $\mathcal{P}$ does not necessarily coincide with the minimum $\chi^2$ value of an individual SED or SB, and that heterogeneous observational uncertainties within the data set can introduce biases into the fitting process. Rather, the likelihood maximisation procedure selects the parameter set $\mathcal{P}$ that yields the most optimal simultaneous reproduction of all data sets included in the fit.

In the fitting, we used the SED data from Fermi-LAT, HAWC, ARGO, and LHAASO, and SB data from LHAASO only. The differences in angular resolution and instrument response among various detectors may introduce systematic uncertainties in any direct comparison of the measured fluxes. In this work, our primary objective is to model the extended emission within a $6^\circ$ radius around Cyg OB2 as observed by LHAASO up to PeV energies. However, emission from the same region has also been detected by other instruments, Fermi-LAT, HAWC, and ARGO, with reported symmetric two-dimensional Gaussian widths of $2^\circ$, $2.13^\circ$, and $1.8^\circ$, respectively, where the Gaussian tails extend beyond the actually observed emission region. To enable a consistent comparison, we rescale the fluxes reported by Fermi-LAT, HAWC, and ARGO using a two-dimensional Gaussian PSF intensity profile, obtaining correction factors of 0.989, 0.981, and 0.996, respectively, while keeping the LHAASO-measured spectrum and surface-brightness profile unchanged. It is, however, evident that this simplifying assumption of a spatially uniform two-dimensional Gaussian emission profile is not strictly valid for the Cygnus cocoon, where the distributions of gas, particle acceleration sites, and particle transport properties are non-uniform. Consequently, this approach may introduce an inherent spatial and spectral inconsistency in the inferred $\gamma$-ray distributions.


In our model, we varied several input parameters, namely the stellar wind velocity ($u_{sw}$), the stellar mass-loss rate ($\dot{M}_{\star}$), the ambient density ($\varrho_0$), the system age ($\mathcal{T}_{\mathrm{age}}$), the spectral index ($s$), the fraction of the wind kinetic energy converted into magnetic energy ($\eta_B$), the fraction of the wind kinetic energy converted into cosmic-ray acceleration ($\epsilon_{\mathrm{cr}}$) and the radial distance of clouds located within the bubble, subject to the constraint that their positions lie between \(R_{sh}\) and \(R_{b}\). For each distinct combination of these parameters, the $\gamma$-ray SED and SB profile are computed with the model and compared with the observational data to identify the parameter set that best reproduces the measurements. The allowed range for the stellar wind velocity ($u_{sw}$) in the fitting procedure is 1000--3000 km s$^{-1}$. Since Cyg OB2 hosts three powerful WR stars, they can drive strong stellar winds and the average wind speed is expected to fall within the range specified in the prior distribution. The total stellar mass-loss rate ($\dot{M}_{\star}$) is varied between $10^{-6}$ and $10^{-3}\,M_\odot\,\mathrm{yr}^{-1}$. For the ambient density of the ISM ($\varrho_0$), we adopt a plausible range of $1-10$ particles cm$^{-3}$. The system age is taken to be $\mathcal{T}_{\mathrm{age}} \in [2,7]$ Myr, the spectral index $s \in [4,5]$, and the fractions of the wind kinetic energy converted into magnetic energy ($\eta_B$) and into cosmic rays ($\eta_{\mathrm{CR}}$) are allowed to vary between 0.1\% and 10\% of the total wind luminosity $\mathcal{L}_{sw}$.



The sets of input parameters $\mathcal{P}$ for the three diffusion types considered in this work, which provide the best joint fit to the observed SED and SB data $\mathcal{D}$ of Cygnus bubble through maximisation of the log-likelihood function $\mathscr{L}$, are summarised in Tab.~\ref{tab:ModelFit}. The analysis is restricted to a predefined region of parameter space, within which the value of the log-likelihood function $\mathscr{L}$ is computed. By employing this type of fitting procedure, we aim to break the degeneracies present within the parameter space. The correlations among parameters within the adopted prior ranges are inferred from the posterior distributions obtained from the fit. These posterior distributions, displayed as corner plots, are shown for the three diffusion types in Figs.~\ref{fig:EBohm}--\ref{fig:EKra} of Appendix~\ref{seca:fitMCMC}.

\begin{figure}
    \centering
    \includegraphics[width=0.95\linewidth]{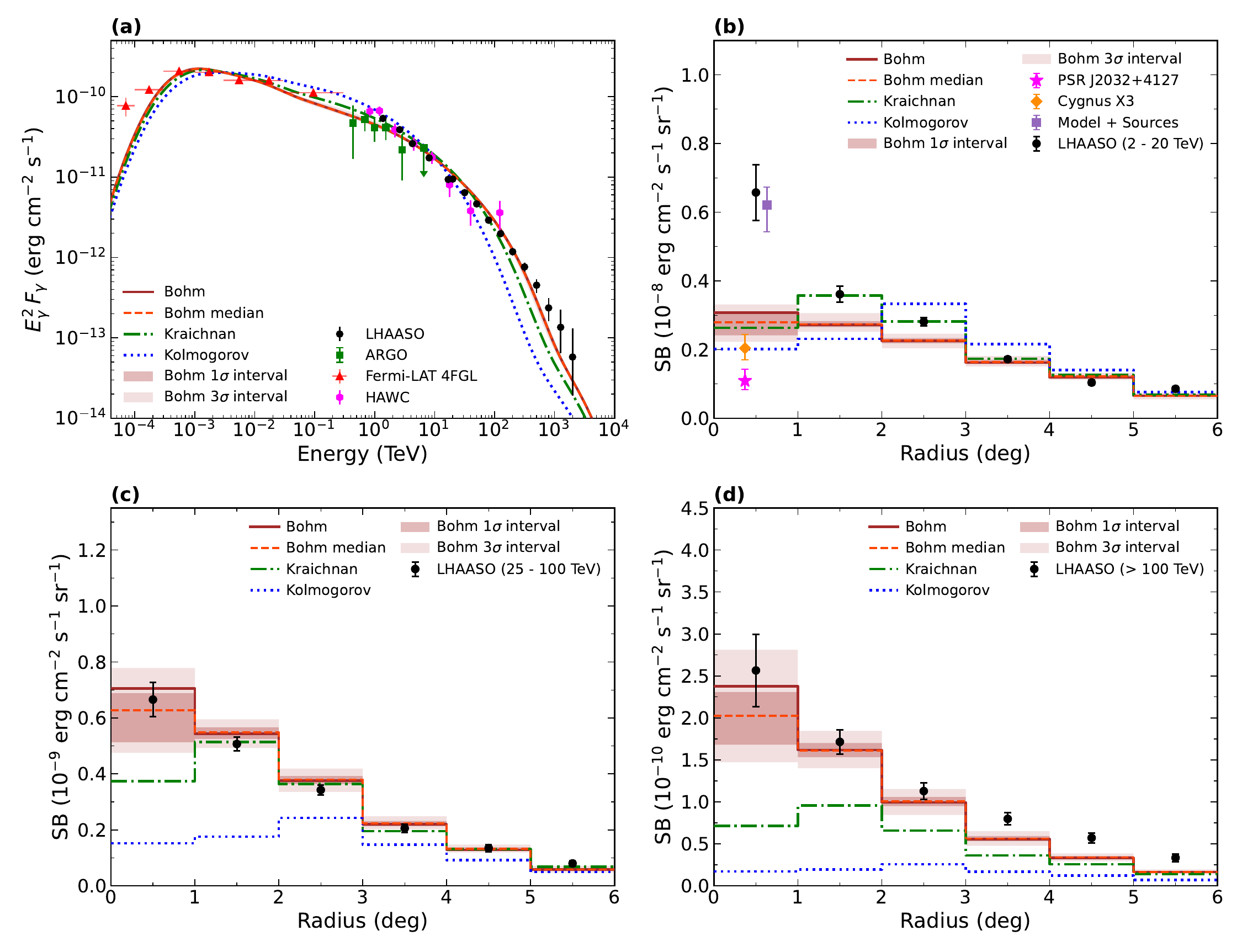}
    \caption{\textbf{(a)} The $\gamma$-ray spectral energy distribution, calculated for the three diffusion models considered in this work, is shown for an angular extension of up to $\sim 6.0^\circ$ of the Cygnus bubble. The curves correspond to the median parameter sets obtained from the joint fitting procedure. The shaded region represents the $1\sigma$ and $3\sigma$ credible interval (which is not observable because it comparable to line widths) of the parameters inferred for Bohm-type diffusion. For comparison, observational data from various $\gamma$-ray instruments are also depicted. The surface brightness profiles of $\gamma$-rays in three energy bands. 
        \textbf{(b)} Low-energy band (2 -- 20 TeV). 
        \textbf{(c)} Mid-energy band (25 -- 100 TeV). 
        \textbf{(d)} High-energy band ($>$ 100 TeV). 
        Each panel shows results for Kraichnan, Kolmogorov, and Bohm-type diffusion. The surface brightness is derived from observations of the centrally located pulsar PSR J2032+4127, and the microquasar Cygnus X-3 is additionally presented for comparison.}
    \label{fig:sed}
\end{figure}

\begin{table*}
    \centering
    \renewcommand{\arraystretch}{1.25}
    \setlength{\tabcolsep}{5pt}
    \resizebox{\textwidth}{!}{%
    \begin{tabular}{llccccc}
        \hline\hline
        \multicolumn{7}{c}{\textbf{MCMC-fit model parameters for the Cygnus bubble}}\\
        \hline
        &\textbf{Parameter} & \multicolumn{3}{c}{\textbf{Best fit}} & \textbf{Fit range} & \textbf{Prior/Range}\\
        && \textbf{Kolmogorov} & \textbf{Kraichnan} & \textbf{Bohm} & Bohm &\\
        \hline
        \multirow{13}{*}{\rotatebox[origin=c]{90}{Fitted}} &
        Wind velocity $u_{\rm sw}$ [km s$^{-1}$] &
            2999.31 &
            2999.96 &
            2961.55 &
            $2896.14^{+98.94}_{-357.29}$ &
            1000--3000\\
        &Mass-loss rate $\dot{M}_\star$ [$M_\odot$ yr$^{-1}$] &
            $9.99\times10^{-4}$ &
            $1\times10^{-3}$ &
            $9.49\times10^{-5}$ &
            $(1.23^{+0.75}_{-0.49})\times10^{-4}$ &
            $10^{-6}$--$10^{-3}$\\
        &Ambient density $\varrho_0$ [$m_p$ cm$^{-3}$] &
            1.02 &
            1.08 &
            6.58 &
            $7.67^{+2.24}_{-3.56}$ &
            1--10\\
        &System age $\mathcal{T}_{\rm age}$ [Myr] &
            5.83 &
            2.40 &
            6.80 &
            $6.69^{+0.29}_{-0.96}$ &
            2--7\\
        &Magnetic-energy fraction $\eta_B$ &
            0.1 &
            0.1 &
            0.078 &
            $0.068^{+0.030}_{-0.032}$ &
            0.001--0.1\\
        &Spectral index $s$ &
            4.17 &
            4.29 &
            4.34 &
            $4.34^{+0.012}_{-0.017}$ &
            4--5\\
        &CR-energy fraction $\epsilon_{\rm cr}$ &
            0.021 &
            0.010 &
            0.020 &
            $0.016^{+0.012}_{-0.005}$ &
            0.001--0.1\\
        &Cygnus North Filament [pc]& 
        171.37& 
        168.96& 
        161.07 
        & 
        $160.49^{+20.31}_{-86.64}$&
        10--260\\
        &Cygnus Central Clumps [pc]& 
        176.06& 
        179.25& 
        30.18
        & 
        $32.43^{+138.51}_{-3.34}$&
        10--260\\
        &Cygnus West [pc]& 
        88.33& 
        166.64& 
        159.44
        & 
        $159.13^{+21.72}_{-88.49}$&
        10--260\\
        &Cygnus SFR South [pc]& 
        86.91& 
        60.04& 
        163.43
        & 
        $162.84^{+10.68}_{-9.35}$&
        10--260\\
        &Cygnus South Filament [pc]& 
        171.28& 
        166.43& 
        169.14
        & 
        $162.96^{+14.11}_{-9.91}$&
        10--260\\
        &Cygnus Far South [pc]& 
        88.02& 
        159.01& 
        87.35
        & 
        $158.13^{+25.18}_{-89.95}$&
        10--260\\
        \hline
        \multirow{5}{*}{\rotatebox[origin=c]{90}{Derived}} &
        Total wind luminosity $\mathcal{L}_{\rm sw}$ [erg s$^{-1}$] &
            $2.83\times10^{39}$ &
            $2.83\times10^{39}$ &
            $2.62\times10^{38}$ &
            --- &
            ---\\
        &Shock radius $\mathcal{R}_{\rm sh}$ [pc] &
            86.84 &
            59.85 &
            26.03 &
            --- &
            ---\\
        &Bubble radius $\mathcal{R}_b$ [pc] &
            394.18 &
            228.73 &
            185.14 &
            --- &
            ---\\
        &Post-shock $\delta \mathcal{B}_{\rm sw}$ [$\mu$G] &
            12.01 &
            17.45 &
            10.85 &
            --- &
            ---\\
        &Maximum energy $E_{\rm max}$ [PeV] &
            14.83 &
            7.63 &
            3.28 &
            --- &
            ---\\
        \hline
        &Combined reduced chi-square $\chi^2_{\rm red}$ &
            31.88 & 10.09 & 6.16 & --- & ---\\
        \hline\hline
    \end{tabular}%
    }
    \caption{The joint-fit parameters of the wind-bubble model for the Cygnus bubble ($\sim6^\circ$). Best-fit values are listed for the Kolmogorov, Kraichnan, and Bohm diffusion cases; for Bohm, we additionally report the 95\% containment band range of posterior mass. The rightmost column indicates the uniform prior ranges adopted for the fitted parameters.}
    \label{tab:ModelFit}
\end{table*}

The results from posterior distributions indicates that, for the Bohm diffusion scenario, the inferred input parameters are largely constrained within the adopted prior bounds. In contrast, for the alternative diffusion types, the likelihood preferentially drives one or more parameters toward the boundary of the prior range, suggesting weaker constraints within the specified parameter space. To determine which diffusion model yields the most accurate representation of the observations, we evaluate the reduced chi-squared statistic, defined as the total chi-squared value divided by the number of degrees of freedom (d.o.f.). On the basis of the best-fit results and the corresponding reduced chi-squared values reported in Tab.~\ref{tab:ModelFit}, we find that the Bohm-type diffusion model provides the most adequate fit to the observational data. A more detailed comparison of the modelling outcomes for the various diffusion prescriptions is provided below. The model-predicted SEDs for the best-fit parameter sets, together with the associated SB profiles, are shown in Fig.~\ref{fig:sed}.

The Kolmogorov-type diffusion scenario is highly inefficient at confining particles, thereby limiting the maximum energy achievable by accelerated particles. Consequently, this diffusion type requires a very high maximum energy to fit the observation. The inferred stellar wind luminosity, $\mathcal{L}_{sw} = 2.83\times10^{39}\ \mathrm{erg\ s^{-1}}$, corresponding to $u_{sw} = 2999.31\ \mathrm{km\ s^{-1}}$ and $\dot{M}_{\star} = 9.99\times10^{-4}\ M_\odot\ \mathrm{yr^{-1}}$, exceeds by approximately an order of magnitude the luminosities reported in the literatures \citep{menchiari2024cygnus}. The corresponding best-fit solution yields the largest reduced chi-squared, $\chi^2_{\mathrm{red}} = 31.88$. Therefore, within the framework of this model, Kolmogorov-type diffusion appears highly unlikely to play a dominant role in particle acceleration in the Cygnus cocoon region. 

Kraichnan-type diffusion provides a modestly improved reproduction of the observed spectral and spatial distributions, yielding a substantially lower reduced chi-squared of $\chi^2_{\mathrm{red}} = 10.09$ for the best-fitting parameter set. In this solution, the parameters $u_{\rm sw} = 2999.96\ \mathrm{km\ s^{-1}}$ and $\eta_B = 10\%$ remain close to the upper bounds of their allowed intervals, implying a mechanical wind luminosity of $\mathcal{L}_{sw} = 2.83\times10^{39}\ \mathrm{erg\ s^{-1}}$ for a mass-loss rate of $\dot{M}_{\star} = 1\times10^{-3}\ M_\odot\ \mathrm{yr^{-1}}$. The same parameter set predicts a downstream magnetic-field perturbation of $\delta \mathcal{B}_{sw} = 17.45\ \mu\mathrm{G}$. Although Kraichnan-type diffusion cannot be conclusively excluded for the Cygnus bubble region, the fit requires extreme values of $\mathcal{L}_{sw}$ and a maximum particle energy of $E_{\max} = 7.63\ \mathrm{PeV}$ to match the observations, conditions that are challenging to realise in stellar clusters of the type considered here. Moreover, while the Kraichnan-type parameter set can reproduce the observed SED, it does not recover the SB profile as a function of LOS radius within the inner core ($\lesssim 2^\circ$). This discrepancy suggests that an alternative diffusion prescription and/or additional particle-injection components are required to provide a self-consistent explanation of the full observed data set.

Finally, in the Bohm-type diffusion case, the best-fitting model yields a reduced chi-squared value of $\chi^2_{\mathrm{red}} = 6.16$, the lowest among all diffusion scenarios examined in this study. The posterior distributions of most parameters are well constrained within the adopted prior ranges, allowing the determination of robust $1\sigma$ and $3\sigma$ credible intervals consistent with the observational data. In this regime, Bohm-like turbulence promotes efficient particle confinement, thereby permitting the highest achievable energies in the accelerated population. The correspondingly suppressed diffusion coefficient increases the residence time of high-energy particles within the system, thereby amplifying the predicted high-energy emission. The best-fit values, $u_{sw} = 2961.55\ \mathrm{km\ s^{-1}}$, $\dot{M}_{\star} = 9.49\times10^{-5}\ M_\odot\ \mathrm{yr^{-1}}$, $\epsilon_{cr} \sim 2\%$ and $\eta_B = 7.8\%$, imply a stellar-wind mechanical power of $\mathcal{L}_{sw} = 2.62\times10^{38}\ \mathrm{erg\ s^{-1}}$, which is markedly lower than that inferred for the other two diffusion types. The corresponding best-fit age of the Cygnus bubble is 6.80 Myr, comparable to the epoch of peak activity of Cyg OB2. 

The collective stellar winds generate a TS region at a radius of $\mathcal{R}_{sh} \simeq 26\ \mathrm{pc}$, while the resulting bubble is predicted to have a radius $\mathcal{R}_{b} > 185\ \mathrm{pc}$, extending well beyond our observational field, which is limited to $\sim 6^\circ \simeq 173\ \mathrm{pc}$ at a distance of 1.65 kpc. Given the structure of the YMSC and the assumption that particle acceleration occurs predominantly at the TS, the SB profile predicted by our model deviates from a simple power law. Specifically, within the inner core region ($l \leq \mathcal{R}_{sh}$), the predicted flux is comparatively suppressed. Although the Bohm-type diffusion reproduces the global observational characteristics well, it slightly underpredicts the very high energy data points in the energy spectrum. This discrepancy is apparent for the SB measurement in the first-bin value of the lowest energy interval (2–20 TeV) and for the outer-edge values in the highest energy interval ($>$100 TeV). 

\begin{figure}
    \centering
    \includegraphics[width=0.65\linewidth]{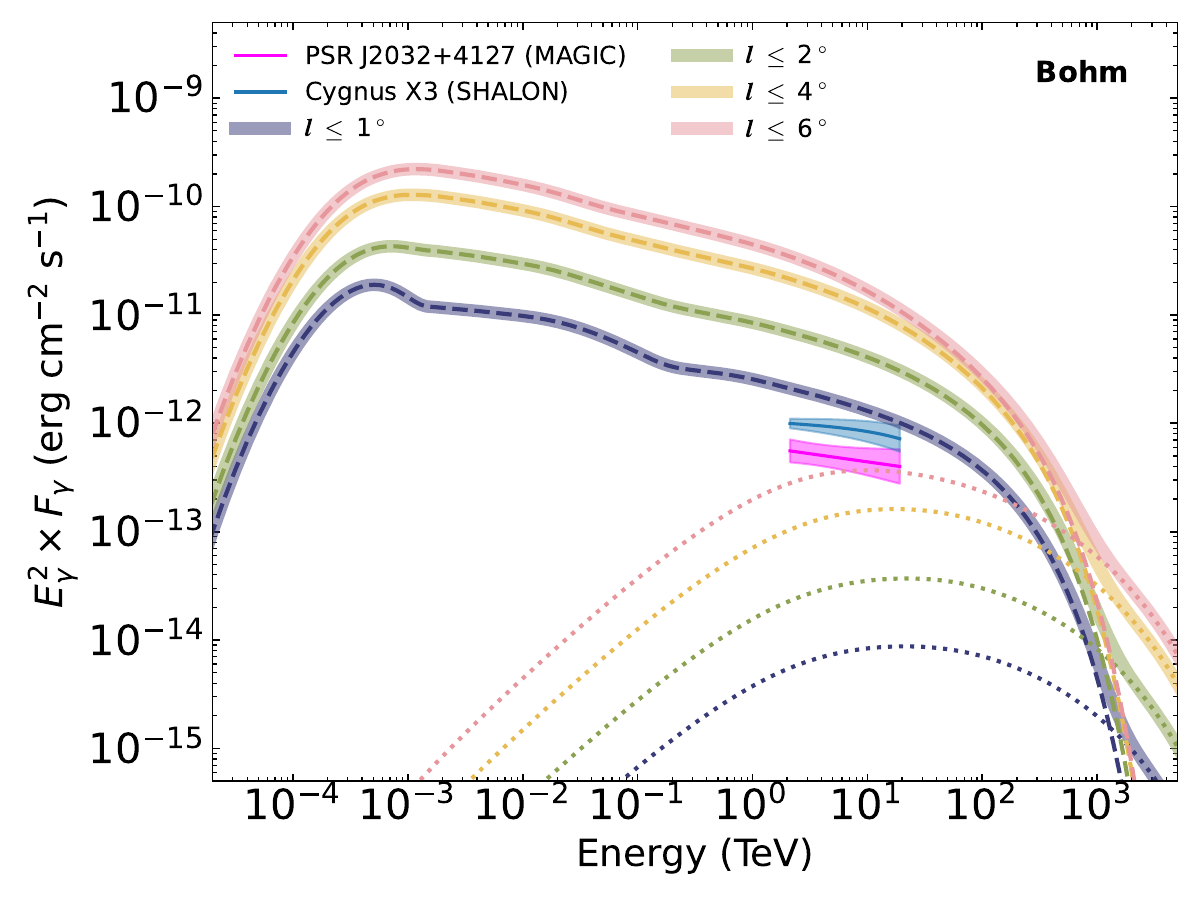}
    \caption{The LOS integrated $\gamma$-ray flux distribution in the Cyg OB2 cluster for Bohm diffusion scenarios is shown. The dashed and dotted lines represent the YMSC and GCR contributions, respectively, for different LOS radii, while the thick continuous line denotes the total contribution within the corresponding radii. For comparison, the $\gamma$-ray emission observed from PSR J2032+4127 and the microquasar Cygnus X-3 with the MAGIC \citep{2018ApJ...867L..19A} and SHALON \citep{2020JPhCS1468a2092S} instruments is also displayed.}
    \label{fig:sed_los}
\end{figure}


Given the presence of additional unresolved sources in the inner-core region, specifically PSR J2032+4127 and the microquasar Cygnus X-3, we infer that the measured emission likely represents a superposition of contributions from the YMSC and these unresolved components. To quantify this, we computed the SB using the best-fitting spectral models to the observed quiescent-state emission of PSR J2032+4127 and Cygnus X-3. The differential energy spectrum of Cygnus X-3 is obtained by fitting SHALON data in the energy range 0.8--100 TeV and is given by $dN/dE = (6.6\pm0.5)\times10^{-13}\ E^{-2.04\pm0.09}$ $\rm exp(-E/(72\pm8)\ TeV)$ $\rm TeV^{-1}\ cm^{-2} s^{-1}$ \citep{2020JPhCS1468a2092S}. The baseline power-law emission of the pulsar PSR J2032+4127 is adopted from \cite{2018ApJ...867L..19A} and is parameterised as $dN/dE = (2.63\pm0.60)\times10^{-14}\ (E/3.50\ \rm TeV)^{-2.06\pm0.17}$ $\rm TeV^{-1}\ cm^{-2} s^{-1}$. For the quiescent state, the $\gamma$-rays corresponding to only the low-energy band of the LHAASO observations are available; in contrast, during active states, the emission extends to higher energies and the flux can increase by approximately an order of magnitude \citep{2024ApJ...975..198A, 2025arXiv251216638T}. The resulting differential energy distributions of these sources are shown in Fig.~\ref{fig:sed_los}, together with the LOS integrated $\gamma$-ray emission at different annuli. The contributions from the YMSC and from GCRs at each radius are represented by dashed and dotted curves, respectively, while the total emission is indicated by the thick solid line. As evident from Fig.~\ref{fig:sed_los}, the GCR contribution in the inner region is very small and is further suppressed at lower energies. This behaviour is expected, since GCRs are assumed to penetrate from outside the bubble, and their density and hence their contribution decreases inside the bubble (see Eqs.~\ref{eq:Nrp_a}-\ref{eq:Nrp_c}). The surface brightness of the two compact sources was then computed using their respective distances (PSR J2032+4127: 1.4 kpc; Cygnus X-3: 7.4 kpc) and renormalised to an angular extent of $1^\circ$. The resulting SB profiles are presented in Fig.~\ref{fig:sed}(b). When the contributions from the pulsar, the microquasar, and the YMSC are combined, the observed SB of the inner core in the 2-20 TeV energy range is satisfactorily reproduced. The $\gamma$-ray emission predicted in the SED, as well as in the outer angular annuli of the SB profile above 100 TeV, however, indicates that, at the highest energies and largest angular separations, an additional emission component is likely required. A plausible origin for this component could be $\gamma$-ray production by particles that escaped from accelerators associated with past explosive events (e.g., $\gamma$-Cygni, SNR G79.8+1.2), or by particle acceleration within the Cygnus stellar cluster itself, followed by subsequent confinement, interactions with foreground and/or background gas reservoirs. A comparable historical outburst, along with the subsequent illumination of escaping particles within the surrounding molecular clouds leading to TeV $\gamma$-ray emission, is studied in \cite{Aharonian_1996emissivity, Gabici_MC_Illumination2009, 2022PASJ...74..625O, Mitchell_Illumination2024lhaaso,Mitchell&Celli24, Oka_2025ApJ...989..137O, 2026arXiv260713991R}. Recently, \cite{2026A&A...712A..41L} concluded that the presence of clouds and the suppression of the diffusion coefficient help explain these emissions.

\begin{figure}
    \centering
    \includegraphics[width=0.65\linewidth]{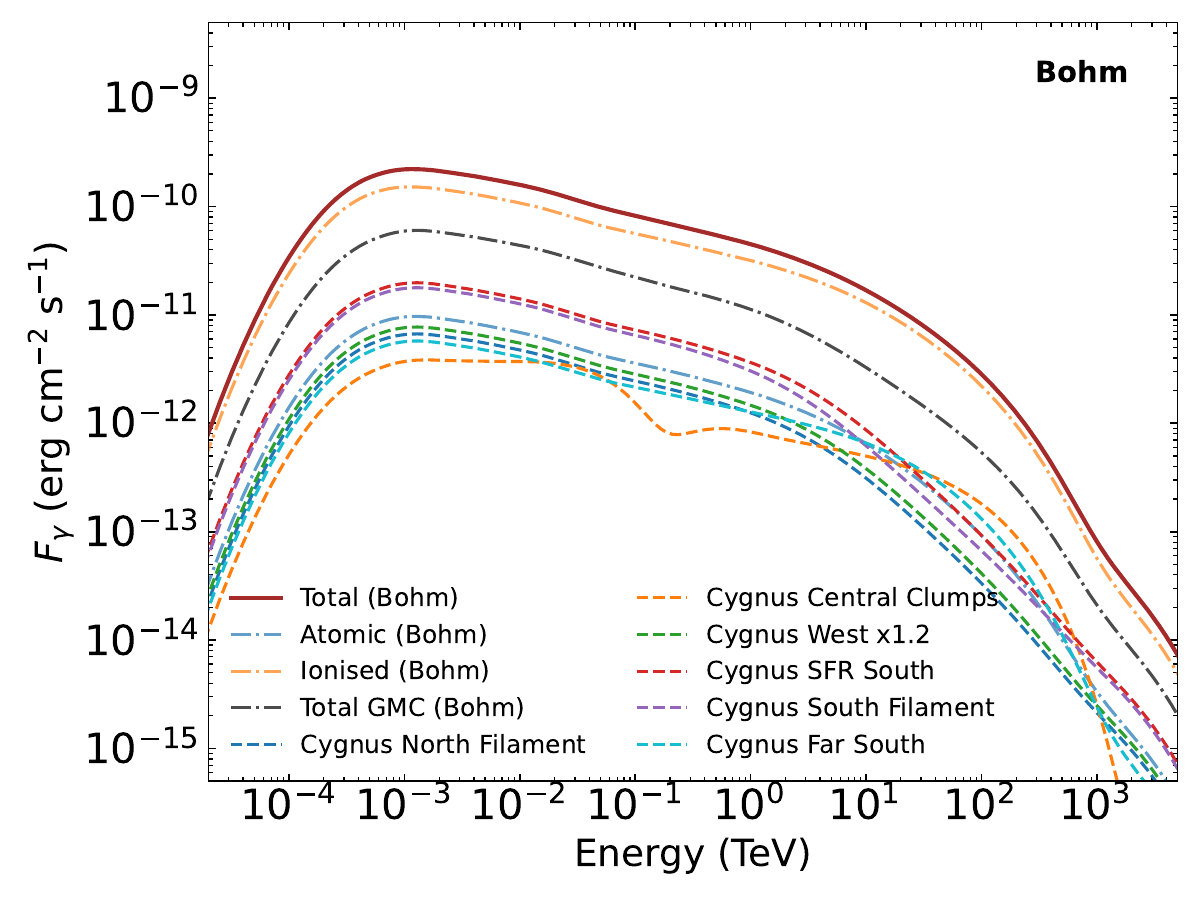}
    \caption{The contribution of individual molecular clouds within the Cygnus bubble to the total $\gamma$-ray emission in the region, under the assumption of Bohm-type diffusion. To enhance visual discriminability, the contribution of the \textit{Cygnus West} cloud was scaled by a factor of 1.2. The total contribution from the atomic gas, ionised gas and the GMCs is also presented for reference.}
    \label{fig:sed_gmcs}
\end{figure}

In Fig.~\ref{fig:sed_gmcs}, the individual contributions of the six GMCs included in the model to the total best-fit observed $\gamma$-ray emission, assuming Bohm-type diffusion, are presented. For comparison, the corresponding contributions from the atomic and ionised gas components are also shown. The figure indicates that the ionised component provides the dominant contribution, followed by the integrated emission from the entire GMC population and by the atomic gas contribution. Differences among the fluxes predicted for the six GMCs are also evident, and are primarily attributable to variations in their physical sizes and their locations within the bubble. In particular, clouds located closer to the shock are expected to exhibit higher fluxes and thus stronger $\gamma$-ray emission. For clouds with otherwise comparable flux conditions, the emission scales with the parameter \textit{A} defined in Sec.~\ref{seca:gasDist}. Consistent with the values of \textit{A} listed in Tab.~\ref{tab:cygnus_clouds}, the southern clouds having larger values are therefore expected to contribute more strongly in the total observed SED and SB profiles. This trend is apparent in Fig.~\ref{fig:sed_gmcs}, especially above 100~TeV, where the integrated surface brightness from these clouds is expected to be higher.

\begin{table*}
\centering
\renewcommand{\arraystretch}{1.25}
\setlength{\tabcolsep}{5pt}
\begin{tabular}{lcccccccccc}
\hline\hline
\multicolumn{10}{c}{\textbf{IceCube}} \\
\hline
\multirow{2}{*}{Diffusion Model} 
    & \multicolumn{3}{c}{1 yr} 
    & \multicolumn{3}{c}{5 yr} 
    & \multicolumn{3}{c}{10 yr} \\
\cline{2-10}
    & $\mathcal{N}_{event}$ & $\mathcal{N}_{bkg}$ & $\sigma$
    & $\mathcal{N}_{event}$ & $\mathcal{N}_{bkg}$ & $\sigma$
    & $\mathcal{N}_{event}$ & $\mathcal{N}_{bkg}$ & $\sigma$ \\
\hline
Bohm        & 9.47 & 53.03 & 1.30  & 47.33 & 265.16 & 2.91 & 94.67 & 530.33 & 4.11 \\
Kraichnan   & 10.27 & 53.03 & 1.41  & 51.33 & 265.16 & 3.15 & 102.65 & 368.69 & 4.46 \\
Kolmogorov  & 10.03 & 53.03 & 1.38  & 50.14 & 265.16 & 3.08 & 100.27 & 368.69 & 4.35 \\
\hline

\multicolumn{10}{c}{\textbf{KM3NeT/ARCA}} \\
\hline
Bohm        &  5.85    &   33.77    &   1.01    &   29.24    &   168.84     &   2.25   &   58.47    &    337.68    &   3.18   \\
Kraichnan   &  6.33    &   33.77    &   1.09    &   31.65    &   168.84     &   2.44   &   63.30    &    337.68    &   3.44   \\
Kolmogorov  &  6.17    &   33.77    &   1.06    &   30.87    &   168.84     &   2.38   &   61.74    &    337.68    &   3.36   \\
\hline\hline
\end{tabular}
\caption{The expected number of neutrino events within the 1 to 1000 TeV energy range originating from the Cygnus cocoon ($\delta ec  = 40.09^\circ$), along with the corresponding detection significance attainable by the IceCube and KM3NeT/ARCA neutrino telescopes, assuming an effective observational period of 1, 5 and 10 years.}
\label{tab:eventRate}
\end{table*}

\subsection{The Detection Significance in IceCube and KM3NeT}
\label{sec:EventRate}
\begin{figure}
    \centering
    \includegraphics[width=0.65\linewidth]{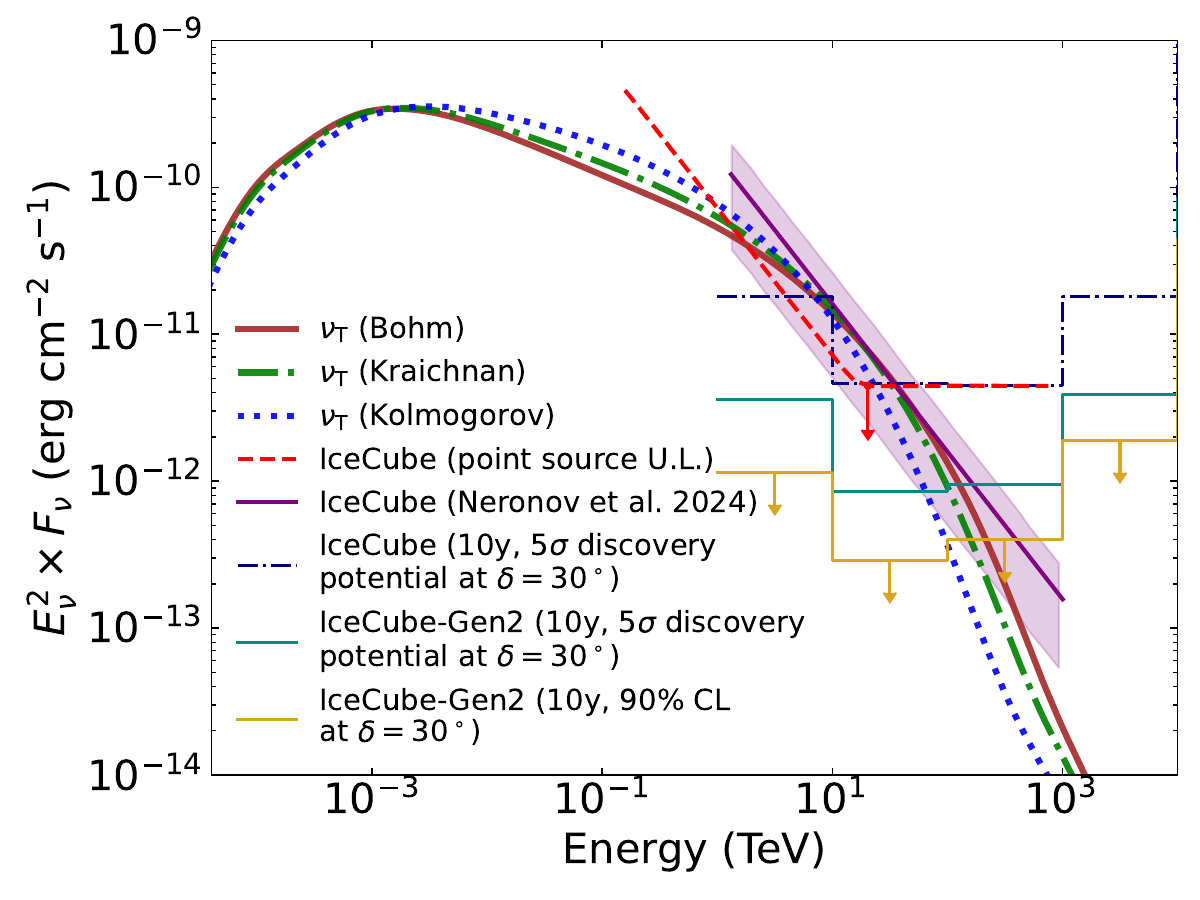}
    \caption{The all flavors ($\nu+\bar{\nu}$) neutrino distribution corresponding to the fitted $\gamma$-ray spectral parameters is presented here. Additionally, the neutrino distribution observed in IceCube is also included for comparison \citep{neronov24}, point source 2HWC J2031+415 upper limit \citep{2020PhRvL.124e1103A} and the sensitivity curves of IceCube detector.}
    \label{fig:nutrinos}
\end{figure}

The model adopted in this work is based on hadronic particle acceleration and is employed to account for the observed $\gamma$-ray emission; the same underlying interactions are also responsible for producing neutrinos. In this section, we present the corresponding neutrino flux predicted by our framework, using the parameter set derived from the spectral and spatial $\gamma$-ray data of the Cygnus bubble. The resulting all-flavour neutrino spectrum is shown in Fig.~\ref{fig:nutrinos}, together with the neutrino flux inferred from 10 years of publicly available IceCube observations. An analysis of muon-neutrino events in the 10-year IceCube data set, in the direction of the Cygnus cocoon, yields only an upper limit for a point-like source at the position of 2HWC J2031+415 \citep{2020PhRvL.124e1103A}, which is compatible with the $\gamma$-ray measurements. The point-source upper limit, indicated by the dashed line with an arrow, has been multiplied by a factor of 3 under the assumption of a standard flavour composition, in order to compare the muon-neutrino flux with the total all-flavour neutrino flux. The solid purple curve and the associated shaded band represent the all-flavour neutrino emission calculated by \cite{neronov24}, who report a $3\sigma$ excess of neutrino events from the direction of the Cygnus bubble that is inconsistent with expectations from background fluctuations.

The neutrino event rates predicted by the model are consistent with the observational data, supporting the hypothesis that the source operates as a hadronic accelerator. We have subsequently computed the all-flavour neutrino event rate from the Cygnus bubble, as well as the atmospheric neutrino background produced by the interactions of GCRs with the Earth’s atmosphere. Atmospheric neutrinos constitute the dominant contribution to the total detected events and therefore represent the primary background for any ground-based neutrino observatory. The detection significance of the signal in the IceCube and KM3NeT detectors, for different diffusion scenarios and exposure times of 1, 5, and 10 years, is evaluated using the standard expression $\mathcal{N}_{\text{event}}/\sqrt{\mathcal{N}_{\text{bkg}}}$ and the corresponding values are listed in Tab. \ref{tab:eventRate}. For Bohm-type diffusion, the detection significances in IceCube for exposure times of 1, 5, and 10 years are 1.30, 2.91, and 4.11, respectively, while for KM3NeT they are 1.01, 2.25, and 3.18. In contrast, the value for 1 year reported for ARGO J2031+4157 (TeV counterpart of the Cygnus Cocoon) in \cite{2024ApJ...969..161G} with a $\sigma$ value of 1.34 for IceCube and 1.1 in KM3NeT/ARCA.
Owing to the position of the Cygnus bubble in the sky and the currently reported detector sensitivities, KM3NeT is predicted to observe fewer events than IceCube, rendering the KM3NeT site comparatively less favourable for the observation of this particular source. Nonetheless, with an exposure of 10 years, the source is expected to be detectable with a significance of \(\sim3\sigma\). Note that sub-degree angular resolution achievable with muon neutrinos can also be used to compare $\gamma$-ray SB and angular distribution of neutrino events to confirm a common origin. Nonetheless, IceCube observations of neutrinos originating from the Galactic plane also reveal a statistically significant excess of events from the direction of the Cygnus region \citep{2023Sci...380.1338I}, which cannot be accounted for solely by interactions between diffuse GCR and the interstellar gas in the Milky Way \citep{2024JCAP...06..074R}. These results corroborate the interpretation of the Cygnus bubble as a hadronic accelerator and a Galactic neutrino emitter. More stringent constraints will be achievable with next-generation instruments and increased statistics, for example, with upgraded observatories such as IceCube-Gen2, and proposed facilities such as HUNT, TRIDENT, NEON. 

\section{Conclusions}
\label{sec:conclusion}
The multi-messenger emission observed from the Cygnus bubble in the energy range from a few GeV up to PeV energies is well reproduced by the particle-acceleration model at the termination shock of the collective stellar wind considered in this work, developed by \cite{Morlino21}. To break parameter degeneracies and to quantify correlations among the input parameters ($u_{\rm sw}, \dot{M}_\star, \varrho_0, \mathcal{T}_{\rm age}, s, \eta_B, \epsilon_{\rm cr}\ \mathrm{and}$ the relative distances of the molecular clouds with respect to the centre of the bubble), we employ a Bayesian joint-fitting module \textsc{Nautilus}. To constrain the hadronic interactions, we also incorporated an updated gas-density distribution of atomic, molecular, and ionised components. We evaluate the model under three different assumptions for the particle diffusion types in order to identify which diffusion type most consistently reproduces the observed data. The diffusion properties depend on the level of magnetic turbulence generated by various MHD instabilities within the wind bubble. In this scenario, a fraction of the total stellar-wind luminosity is converted into magnetic energy, which in turn regulates the maximum energy that can be attained by accelerated particles.

Among the turbulence regimes considered, Kolmogorov-type diffusion is strongly disfavoured, yielding the largest reduced chi-square and thus providing the least acceptable fit to the observed data. Kraichnan-type diffusion can reproduce the observations reasonably well; however, it requires parameter values that appear difficult to achieve for the type of stellar cluster studied here, namely a wind luminosity of $\mathcal{L}_{\rm sw} = 2.83\times10^{39}\ \mathrm{erg\ s^{-1}}$, a magnetic-energy fraction of $\eta_B = 10\%$, and a maximum particle energy of $E_{\max} = 7.63\ \mathrm{PeV}$. In contrast, the Bohm-type diffusion scenario yields the minimum reduced $\chi^{2}$ and thus provides an adequate description of the observational data. The $\gamma$-ray spatial morphology, quantified via the surface-brightness profile, is also well reproduced by the Bohm-diffusion model over the investigated energy range. The only notable discrepancies are confined to a single inner-core data point in the 2--20 TeV band, as well as a slight underprediction of the $>100$ TeV emission SED and of the SB at the outer edge in the $>100$ TeV band. We show that this discrepancy can be resolved by including additional emission from the pulsar and the microquasar located near the centre of the bubble, in which case the outlying data point is naturally accommodated. The contribution of diffuse GCRs penetrating the bubble cavity from the interstellar medium is significant only at the highest energies (being dominant above 1 PeV); toward the inner core region, this contribution decreases consistently. The neutrino flux distribution and the event rates measured by the IceCube detector are also compatible with this picture, making a predominantly leptonic origin of the emission less favourable, as discussed in Appendix \ref{seca:Lepton}. 

The best-fit system age inferred under the assumption of Bohm-type diffusion is 6.58 Myr, with 95\% posterior mass lying within $6.69^{+0.29}_{-0.96}$ Myr. Within this timescale, at least one type-Ic supernova explosion is expected to have occurred \citep{haerer2025deciphering}. Such an event could modify the particle-acceleration scenario, with an interaction between the SNR and the pre-existing stellar-wind bubble potentially contributing to the observed highest-energy emission, in the regime where the current model slightly underpredicts the data. Nevertheless, as discussed in \cite{Sushch_2025}, although the Hillas criterion provides an upper bound on the achievable particle energy, accelerating particles to PeV energies in such an SNR still remains challenging as the YMSC unless the magnetic field is substantially amplified, thereby enabling Bohm-like diffusion. As shown by \cite{haerer2025deciphering}, the existence of a collective termination shock is not guaranteed when the cluster is too loose. In such a situation, the acceleration model used here could still be applied to the termination shock of every single stellar wind. The total energy budget would remain the same, hence the spectrum and the acceleration efficiency predicted would not change. What, instead, changes significantly is the maximum energy, which is expected to be lower by some factor for the Bohm, Kraichnan or Kolmogorov case, respectively. This would imply that the Bohm case can still provide a viable explanation for the $\gamma$-ray data, with the possible exception of the highest-energy data point, while the other two configurations are strongly disfavoured. Although the assumption of Bohm diffusion throughout the entire bubble is likely unrealistic and should be regarded as an upper-limit case, it remains plausible that an intermediate diffusion (between Kraichnan and Bohm) could also reproduce the observational constraints; however, such parameterizations were not explored in the present study. In such a scenario, $\gamma$-rays beyond $\sim100$ TeV require an alternative explanation. Alternatively, the pulsar, SNR, or microquasar situated near the centre of the bubble may release additional high-energy particles during its active phases. The diffusion of these particles could help to even out the emission throughout the entire cavity and might contribute to the highest-energy emissions observed. However, these possibilities are not considered further here because of the absence of direct observational evidence of an SNR relic in the region, despite the presence of indirect indications such as a pulsar \citep{2015MNRAS.451..581L} and a very massive runaway star \citep{2007A&A...467L..23C} in the Cyg OB2 association. A more detailed exploration of this scenario will be possible in the future, once more definitive observational signatures become available, but such an investigation lies beyond the scope of the present work.

\subsection*{\textit{Acknowledgements}}
{This research made use of the data from the Milky Way Imaging Scroll Painting (MWISP) project, which is a multi-line survey in 12CO/13CO/C18O along the northern galactic plane with PMO-13.7m telescope. We are grateful to all the members of the MWISP working group, particularly the staff members at PMO-13.7m telescope, for their long-term support. MWISP was sponsored by National Key R\&D Program of China with grants 2023YFA1608000 \& 2017YFA0402701 and by CAS Key Research Program of Frontier Sciences with grant QYZDJ-SSW-SLH047. The Canadian Galactic plane Survey is a Canadian project with international partners, and is supported by the Natural Sciences and Engineering Research Council (NSERC). GM acknowledges support from the INAF Theory Grant 2024 {\it Star Clusters As Cosmic Ray Factories II} and European Union – NextGenerationEU RRF M4C2 1.1 under grant PRIN-MUR 2022TJW4EJ. SR acknowledges support by a NITheCS grant from the National Research Foundation of South Africa. We thank R. Aloisio, P. Blasi, and B. Li for reading and giving their feedback. We also thank Cong Li for providing the LHAASO, $\gamma$-ray contour maps.}

\bibliography{ref_prop}
\bibliographystyle{JHEP}

\appendix

\counterwithin{figure}{section}
\section{APPENDIX}
\subsection{The Gas \& Cloud Distribution}
\label{seca:gasDist}
The molecular clouds considered in the vicinity of the Cygnus bubble are associated with a structure at a distance of $\sim 1.65$ kpc, with its centre located at Galactic coordinates $l \sim 80.22039^\circ$ and $b \sim 0.79661^\circ$. The corresponding cloud sample is summarised in Tab.~\ref{tab:cygnus_clouds}. The clouds listed in this table adopt the same baseline properties as in \cite{haerer2025deciphering}. The clouds located in the foreground $\leq1.4$ kpc are excluded. In addition, two columns are included: \textit{A} and $d_{\rm 3D}$. The parameter $A=(M/10^5\,M_\odot)/(d/1\,\mathrm{kpc})$ is derived from the adopted cloud mass and heliocentric distance reported by \cite{Vieu_H_2025}. The three-dimensional separation $d_{\rm 3D}$ is computed as the Euclidean norm of the difference between Cartesian position vectors obtained from the Galactic coordinates via
$x = D \cos b \cos l$, $y = D \cos b \sin l$, and $z = D \sin b$,
assuming the bubble centre is located at $(l,b,D)=(80.22^\circ, 0.80^\circ, 1.65\,\mathrm{pc})$. A modified version of this estimate that incorporates a 5\% uncertainty in the inferred cloud distance is also presented. This shows that for \textit{Cygnus Central Clumps}, located near the central region, the estimate is most strongly affected.


\begin{table}[htbp]
\centering
\small
\caption{Molecular clouds towards the Cygnus region, listing their Galactic longitude ranges and adopted distances (\url{https://doi.org/10.17617/3.G9LUS1}). Adapted from \citep{haerer2025deciphering}; the distance references are those given therein.}
\label{tab:cygnus_clouds}
\begin{tabular}{llccllcc}
\hline\hline
ID & Cloud & Longitude range [$^\circ$] & Distance [kpc] & $A$ & $d_{\rm 3D}$ [pc] & Reference \\
\hline
A & Foreground Rift         & 74--86     & 0.7--1.1  & 10.68 & $650^{+50}_{-49}$ & \citep{2024AJ....167..220Z}\\
B & Cygnus North-West       & 79.5--81   & $\sim$1   & 0.17  & $651^{+50}_{-49}$ & \citep{2024AJ....167..220Z}\\
C & Cygnus 1.3 North        & 80--83.5   & 1.3--1.4  & 0.48  & $352^{+65}_{-63}$ & \citep{2024AJ....167..220Z}\\
D & Cygnus 1.3 South        & 77--80     & 1.2--1.3  & 0.36  & $356^{+63}_{-63}$ & \citep{2024AJ....167..220Z}\\
E & Cygnus North Filament   & 80.5--82   & 1.4--1.5  & 0.28 & $189^{+71}_{-71}$  & \citep{Rygl12}\\
F & Cygnus Central Clumps   & 79--81     & 1.6--1.8 & 0.15 & $87^{+87}_{-58}$    & \citep{2021MNRAS.502.6080O}; \citep{2013ApJ...769...15X}\\
G & Cygnus West             & 78--80     & $>$1.4   & 0.27 & $106^{+64}_{-13}$ & --- \\
H & Cygnus SFR South        & 77--80     & $>$1.4   & 0.83 & $95^{+77}_{-31}$  & ---\\
I & Cygnus South Filament   & 77--79     & $>$1.4   & 0.75 & $92^{+80}_{-37}$  & ---\\
J & Cygnus Far South        & 76--78     & 1.6--1.8 & 0.24 & $106^{+63}_{-14}$ & \citep{Schneider07} \\
\hline\hline
\end{tabular}
\end{table}

\begin{figure}[!ht]
    \centering
    \includegraphics[width=\linewidth]{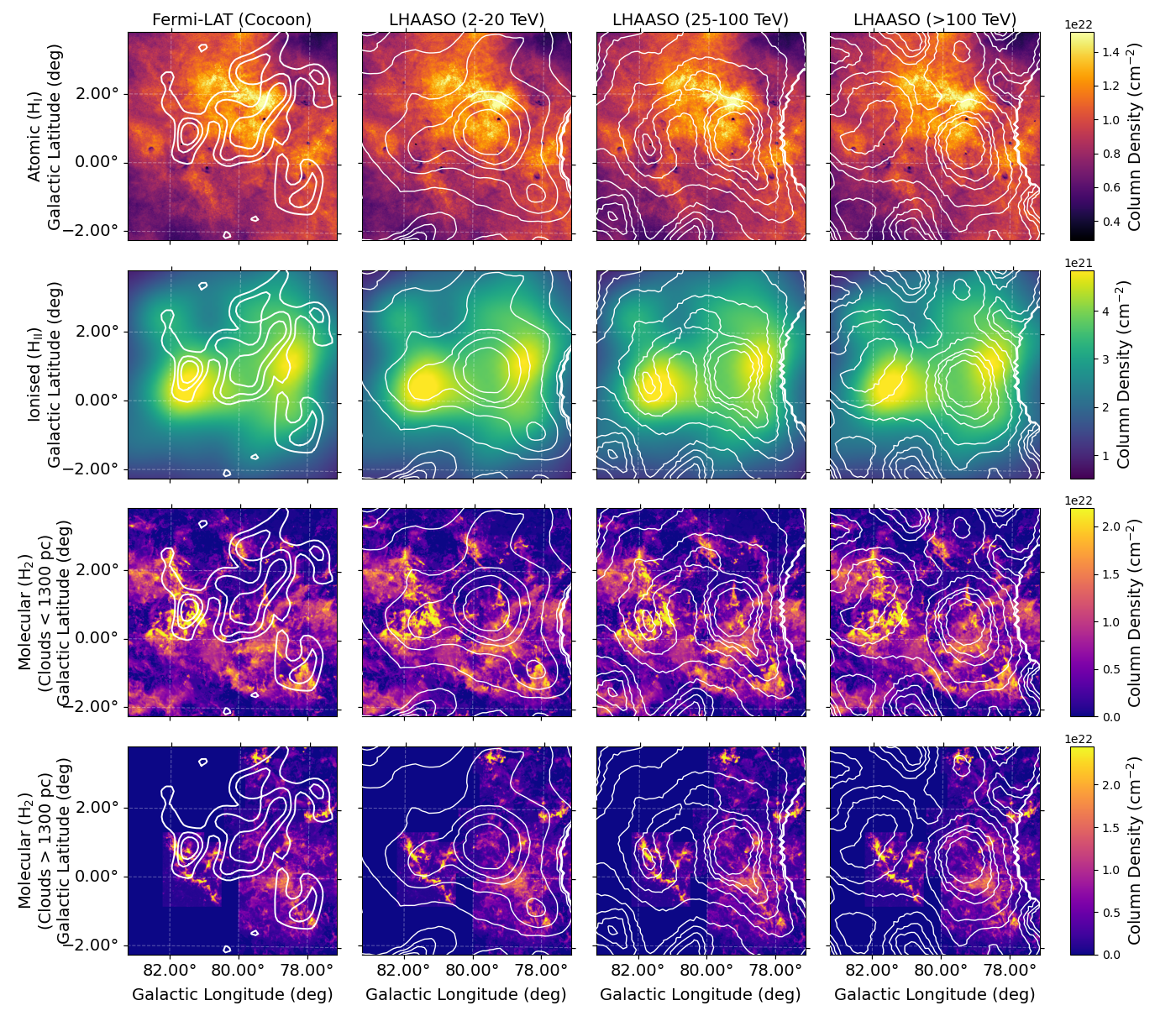}
    \caption{This figure presents the spatial distribution of atomic, ionised, and foreground/background molecular gas in the Cygnus OB2 region. The overlaid contours represent emission measured by the Fermi-LAT and the LHAASO instrument across different energy bands. The colour scale indicates the line-of-sight–integrated column density of the observed gas.}
    \label{fig:2Dgas}
\end{figure}

If the $\gamma$-ray emission detected in the GeV--TeV band arises predominantly from hadronic $p$--$p$ interactions, a strong spatial correlation with the ambient gas density is expected. To test for a common origin, we compared the two-dimensional $\gamma$-ray surface-brightness distributions inferred from Fermi-LAT and LHAASO data with the line-of-sight column densities of atomic, ionised, and molecular gas within the bubble region. We find a pronounced spatial correspondence between the molecular gas component and the $\gamma$-ray emission (Fig.~\ref{fig:2Dgas}). This association supports a hadronic scenario in which accelerated cosmic rays interact with the surrounding medium and generate $\gamma$-rays through inelastic collisions.

\begin{figure*}[!ht]
    \centering
    \includegraphics[width=\textwidth]{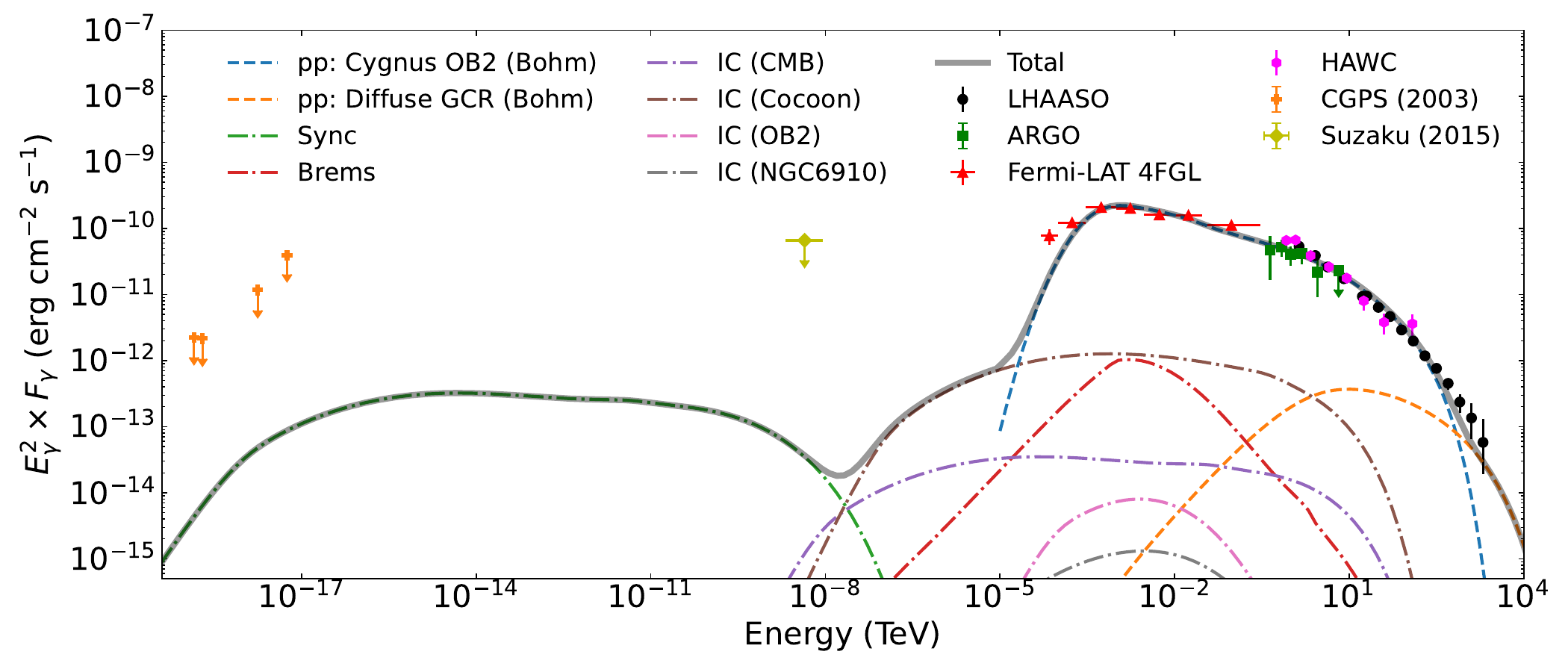}
    \caption{The spectral distribution combining all model components, including the hadronic contribution corresponding to the median parameters of the jointly fitted SED and SB data, is shown under the assumption of Bohm-type diffusion. The individual contributions from hadronic emission associated with the YMSC and from GCRs are presented separately. The leptonic emission arising from secondary leptons produced in hadronic interactions is also displayed, scaled by an appropriate factor. In addition, a possible primary leptonic contribution from the YMSC is shown, assuming a fiducial electron-to-proton normalisation ratio of $10^{-3}$.}
    \label{fig:Total_SED}
\end{figure*}

\begin{figure}[!ht]
    \centering
    \includegraphics[width=0.45\linewidth]{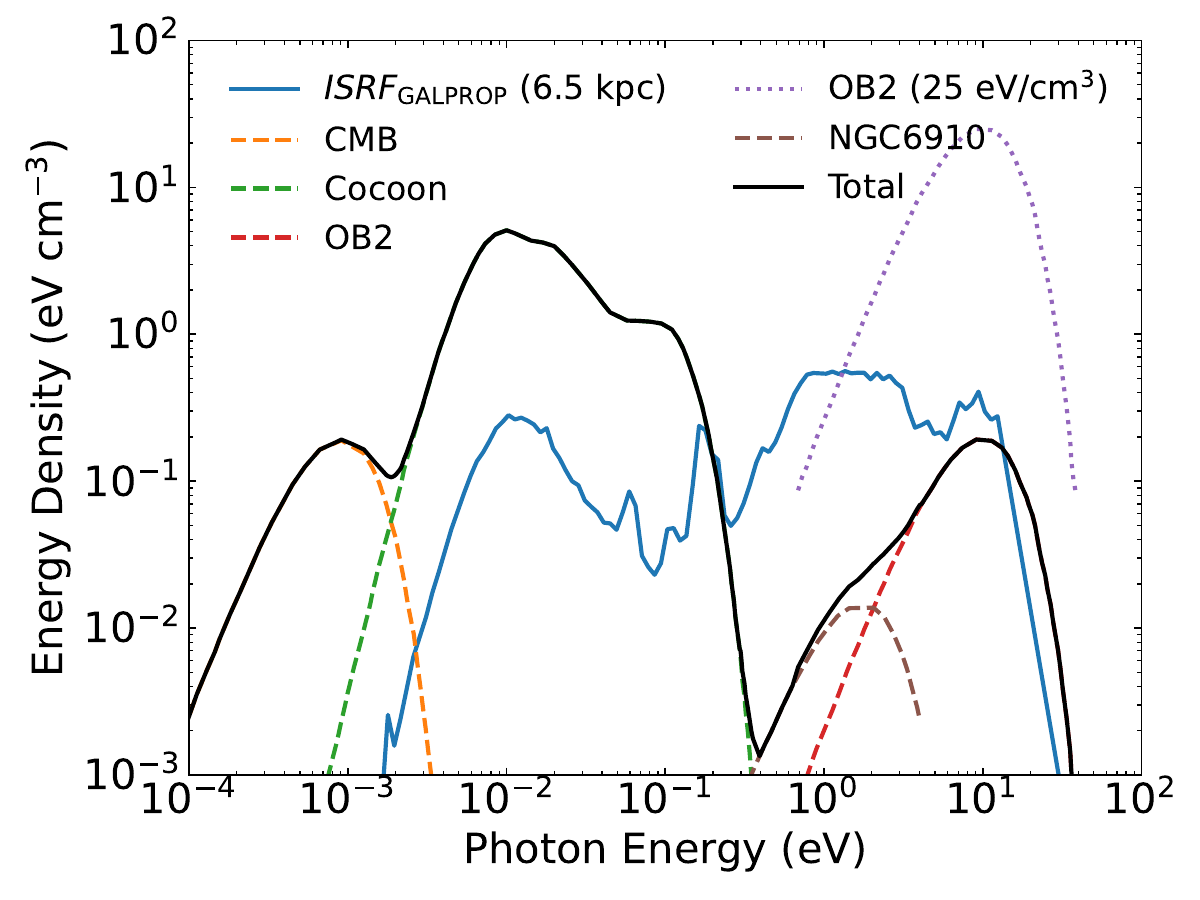}
    \includegraphics[width=0.45\linewidth]{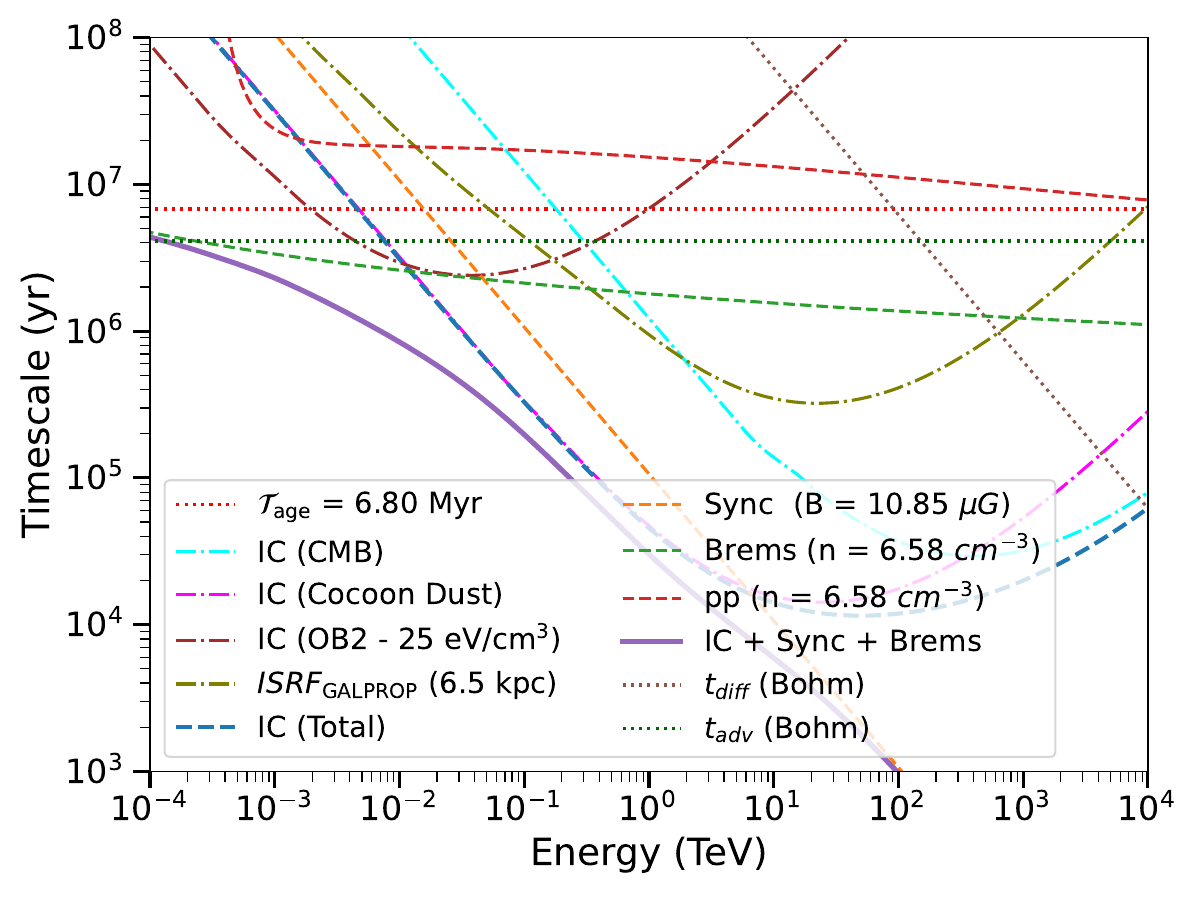}
    \caption{(left) The total density of low-energy photons in the Cygnus bubble region, along with the individual components, is shown here \cite{2011Sci...334.1103A}; for comparison, the soft-photon density at a distance of 6.5 kpc and the enhanced ultraviolet photon field associated with the OB2 cluster are also included. (right) Displayed are the cooling timescales for the various electron and proton interaction processes relevant to the star cluster, computed for the parameter set obtained under the assumption of Bohm-type diffusion (see Tab. \ref{tab:ModelFit}).}
    \label{fig:timescale1}
\end{figure}
\begin{figure}[!ht]
    \centering
    \includegraphics[width=0.45\linewidth]{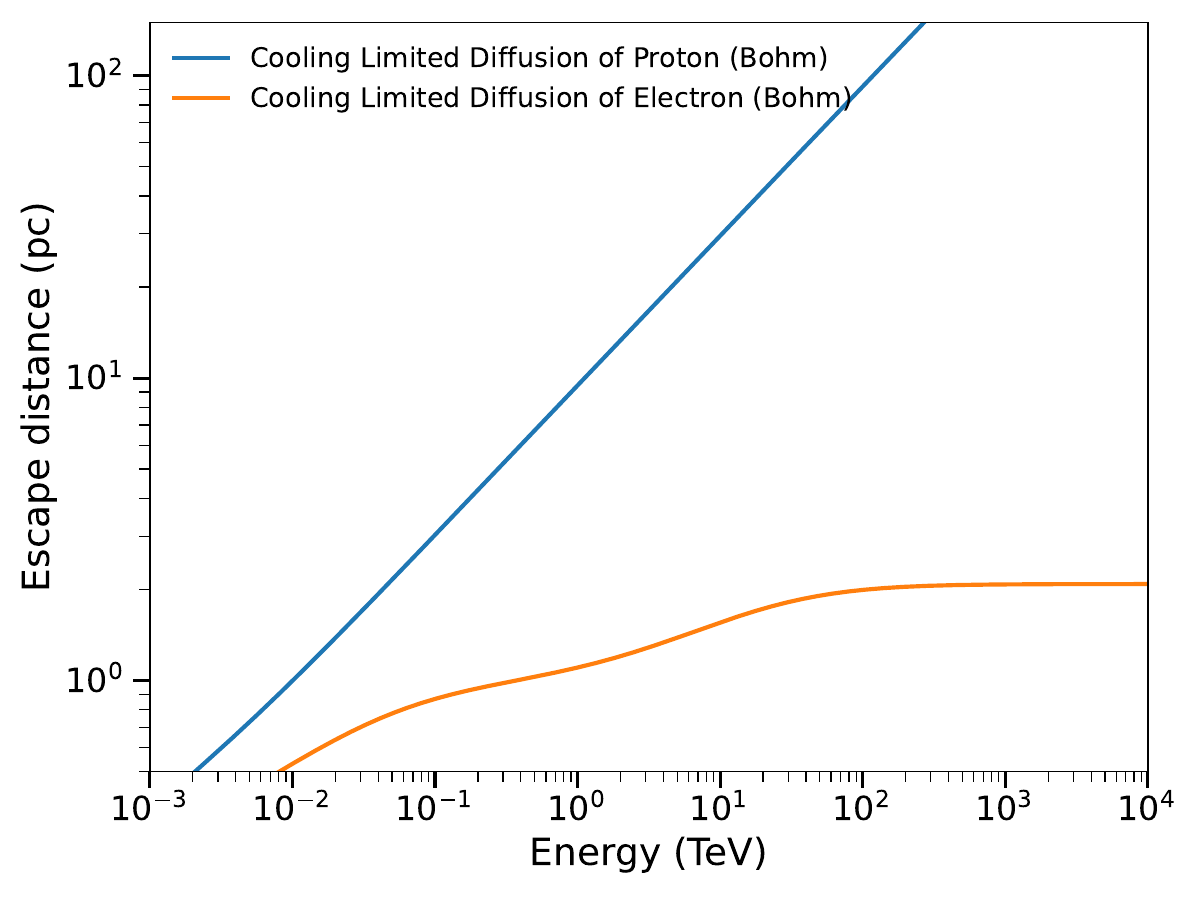}
    \includegraphics[width=0.45\linewidth]{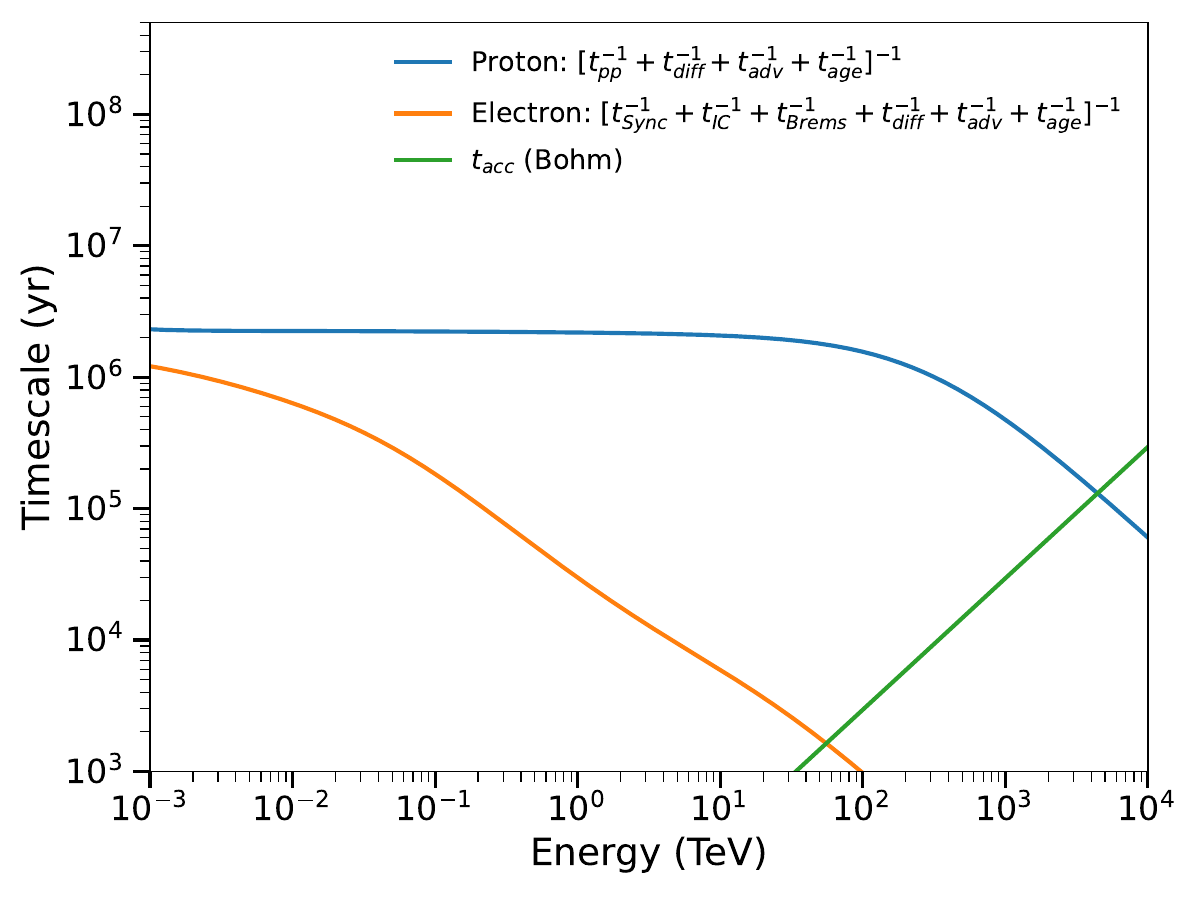}
    \caption{(left) The escape distance of protons and electrons is constrained by their respective cooling timescales within the bubble. (right) The maximum attainable particle energy in the Cygnus cluster for the adopted parameter set, assuming Bohm-type diffusion, is shown. This maximum energy is determined by the intersection of the energy-loss curves with the acceleration timescale.}
    \label{fig:timescale2}
\end{figure}

\subsection{The Leptonic Contribution}
\label{seca:Lepton}
Based on the discussion in Section \ref{sec:EventRate}, the contribution from leptons at energies \(E > 1\ \text{TeV}\) is expected to be minimal or even negligible. Nevertheless, for completeness, we here proceed to estimate such a contribution. All relevant leptonic radiative channels are included, namely synchrotron radiation from electrons in the downstream wind, bremsstrahlung emission produced through interactions with the ambient gas in the cluster medium, and inverse Compton scattering of electrons off the photon fields permeating the cluster environment.


For the synchrotron component, we adopt a downstream magnetic field strength of \textcolor{blue}{10.85} $\mu$G. The gas density is taken from observational estimates, as shown in Fig. \ref{fig:density}. For the target radiation fields, we consider the energy density of the Cosmic Microwave Background (CMB), the infrared emission from dust in the cocoon, and the radiation from the two dominant stellar populations associated with the main clusters Cyg OB2 and NGC 6910, as characterised by \cite{2011Sci...334.1103A} as shown in Fig.~\ref{fig:timescale1}(left). For reference, we also present the Interstellar Radiation Field (ISRF) at a Galactocentric distance of 6.5 kpc as derived from GALPROP \cite{2017ApJ...846...67P}, together with an enhanced ultraviolet (UV) radiation field expected in the vicinity of the stellar core. The comparison indicates that, according to \cite{2011Sci...334.1103A}, the dust emission from this region is substantial.

The electron distribution is modelled following the formalism described by \cite{Peron25}. It is assumed to have the same spectral shape as the proton distribution presented in that work, differing only in its normalisation and maximum energy:

\begin{equation}
    N_{\rm sh}^{e}(p) = k_{\rm ep}\;N_{\rm sh}^{p} (p)\;\exp\left(-\frac{p}{p_{\rm max}^{e}}\right)
\end{equation}
where $k_{ep}$ is the normalization factor, which typically ranges from $10^{-4}$ to a few times $10^{-2}$ in standard SNR scenarios \citep{2021MNRAS.508.6142M, 2017MNRAS.464.2326S, 2015PhRvL.114h5003P}. The electron maximum momentum, $p_{max}^{e}$, is determined by the balance between the acceleration timescale and the relevant radiative loss timescales.

The energy-loss timescales associated with synchrotron radiation ($t_{\rm sync}$), bremsstrahlung ($t_{\rm brems}$), inverse Compton scattering ($t_{\rm IC}$), and proton-proton ($p-p$) ($t_{\rm pp}$) interactions are computed and shown in Fig. \ref{fig:timescale1}(right). For comparison, the timescales for advection ($t_{\rm adv}$), diffusion ($t_{\rm diff}$), and the system age ($t_{\rm age}$), corresponding to the parameter set inferred from fitting the observed $\gamma$-ray data with a Bohm-type diffusion model, are also displayed in the same figure. The figure indicates that energy losses of electrons due to synchrotron radiation and inverse Compton scattering are highly efficient, thereby imposing a strong constraint on the maximum energy that leptons can attain within the Cygnus bubble. It is also evident that, at high energies, synchrotron energy losses and IC energy losses associated with dust emission dominate relative to those predicted under the assumed enhanced UV field and the GALPROP ISRF model, thereby constraining the maximum attainable electron energies. The maximum attainable particle energy in the cluster for this parameter set is determined by solving for the energy at which the acceleration timescale equals the total escape and radiative (or collisional) loss timescales. Specifically, this is obtained by equating the acceleration timescale $t_{\rm acc}$ with the sum of the escape timescale, $1/t_{\rm esc} = 1/t_{\rm adv} + 1/t_{\rm diff} + 1/t_{\rm age}$, and the relevant loss timescale, $1/t_{\rm loss}$, where for leptonic particles $1/t_{\rm loss} = 1/t_{\rm sync} + 1/t_{\rm brems} + 1/t_{\rm IC}$ and for hadronic particles $1/t_{\rm loss} = 1/t_{\rm pp}$.  A comparison of the total loss time scales of protons and electrons with their respective acceleration time scales in Fig.~\ref{fig:timescale2}(right) shows that, on average, the maximum energy achievable by electrons will be in the order of tens of TeV range and is smaller than that of protons due to their cooling at higher energies. The left panel of the figure shows the diffusion length, \(\mathcal{R}_{\rm diff}=\sqrt{4\mathcal{K}_b t_{\rm loss}}\), indicating that, within their respective cooling times in the bubble, electrons exhibit a shorter diffusion length than protons from where they are accelerated.

The same radial distribution adopted for protons is also employed for leptons, but rescaled by the relevant loss and escape time scales in order to account for energy losses during propagation.

\begin{equation}
    N^{e}(r, p) = k_{\rm ep}\,N^{p}(r, p)\,\exp\left(-\frac{p}{p_{\mathrm{max}}^{e}}\right)\,\frac{(t_{\mathrm{esc}}^{-1} + t_{\mathrm{loss}}^{-1})^{-1}_e}{(t_{\mathrm{esc}}^{-1} + t_{\mathrm{loss}}^{-1})^{-1}_p}
\end{equation}

Here, we restrict our analysis to only Bohm-type particle diffusion, since this type of diffusion explains the observation with the minimum reduced chi-square value. We then estimate the total leptonic contribution to the $\gamma$-ray emission by adopting a fiducial normalization factor of $k_{\rm ep} = 10^{-3}$, rather than trying to fit the data. The spatial distribution of electrons is assumed to follow that of the protons; however, energy losses of electrons within the cluster are incorporated by rescaling the spatial profile according to the ratio of the characteristic timescales of protons to electrons. It should be emphasized that our objective here is solely to quantify the expected leptonic contribution, rather than to perform a full lepto-hadronic fit to the data. The combined contributions from all relevant emission processes of both electrons and protons are presented in Fig. \ref{fig:Total_SED}. From the figure, one can see that the leptonic contribution at GeV energies is $\sim2$ orders of magnitude below the hadronic one. Hence, leptonic processes can dominate the emission only if $k_{\rm ep} \gtrsim 0.1$. Such a large value has never been inferred in SNR shocks, nor from observations neither from numerical simulations: $k_{\rm ep}$ typically ranges between $10^{-4}$ and $10^{-2}$. Within the accuracy of this approximate treatment, we therefore disfavour any significant leptonic contribution from the Cyg OB2 to the observed emission if the acceleration is mainly due to stellar winds, contrary to the claim of \cite{haerer2025deciphering}. The Swift-XRT observations provide further support for this hypothesis \citep{2023ApJ...950..116G}.


\subsection{The Bayesian Fitting}
\label{seca:fitMCMC}
\begin{figure}[!ht]
    \centering
    \includegraphics[width=\linewidth]{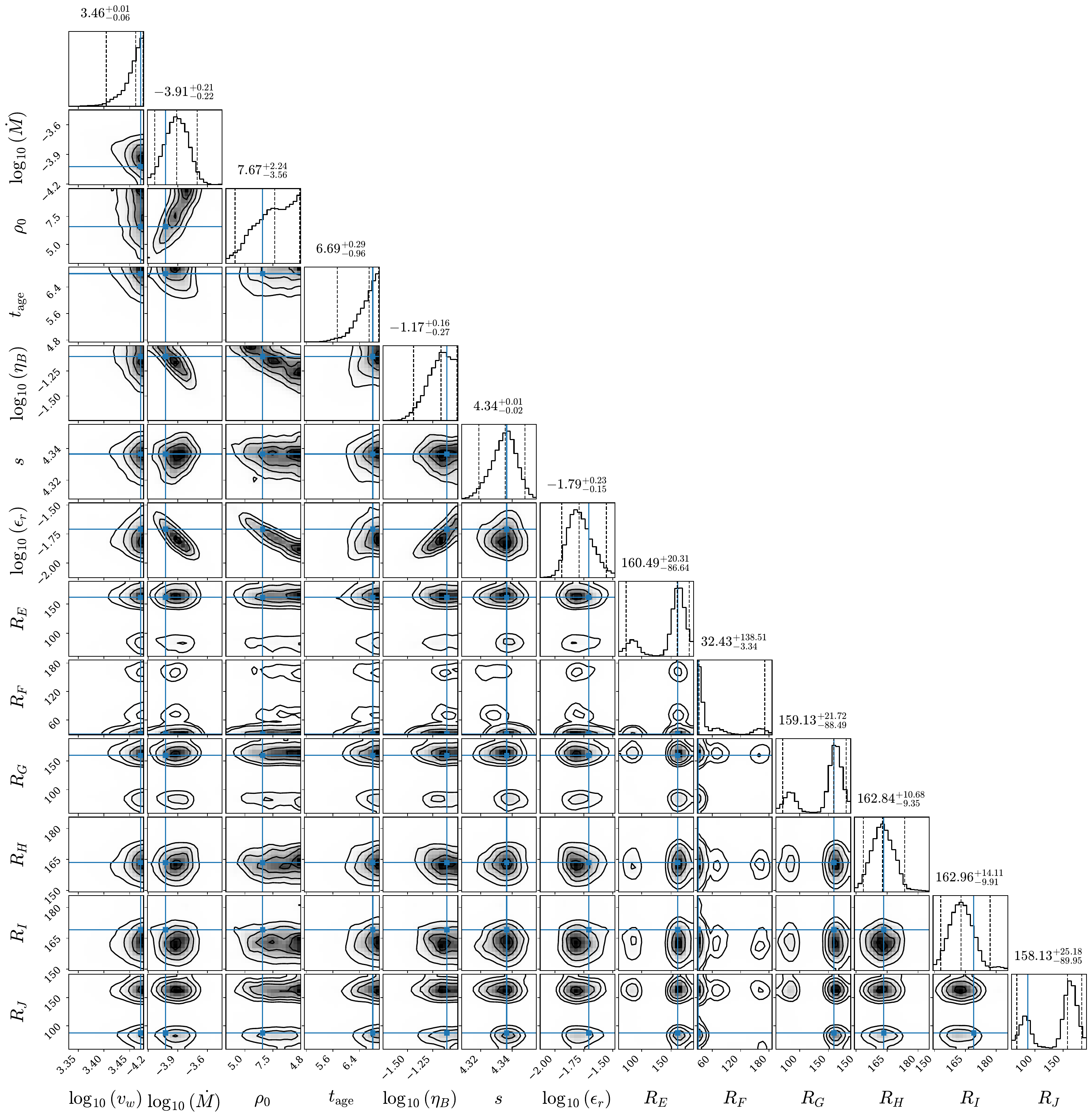}
    \caption{The posterior distribution of the maximised log-likelihood, $\ln \mathscr{L}_i(\mathcal{D}_i \mid \mathcal{P})$, associated with both the spectral and surface brightness profiles of Cyg OB2, derived under the assumption of Bohm-type particle diffusion. The blue lines denote the best-fit of the inferred parameter distribution.}
    \label{fig:EBohm}
\end{figure}

\begin{figure}[!ht]
    \centering
    \includegraphics[width=\linewidth]{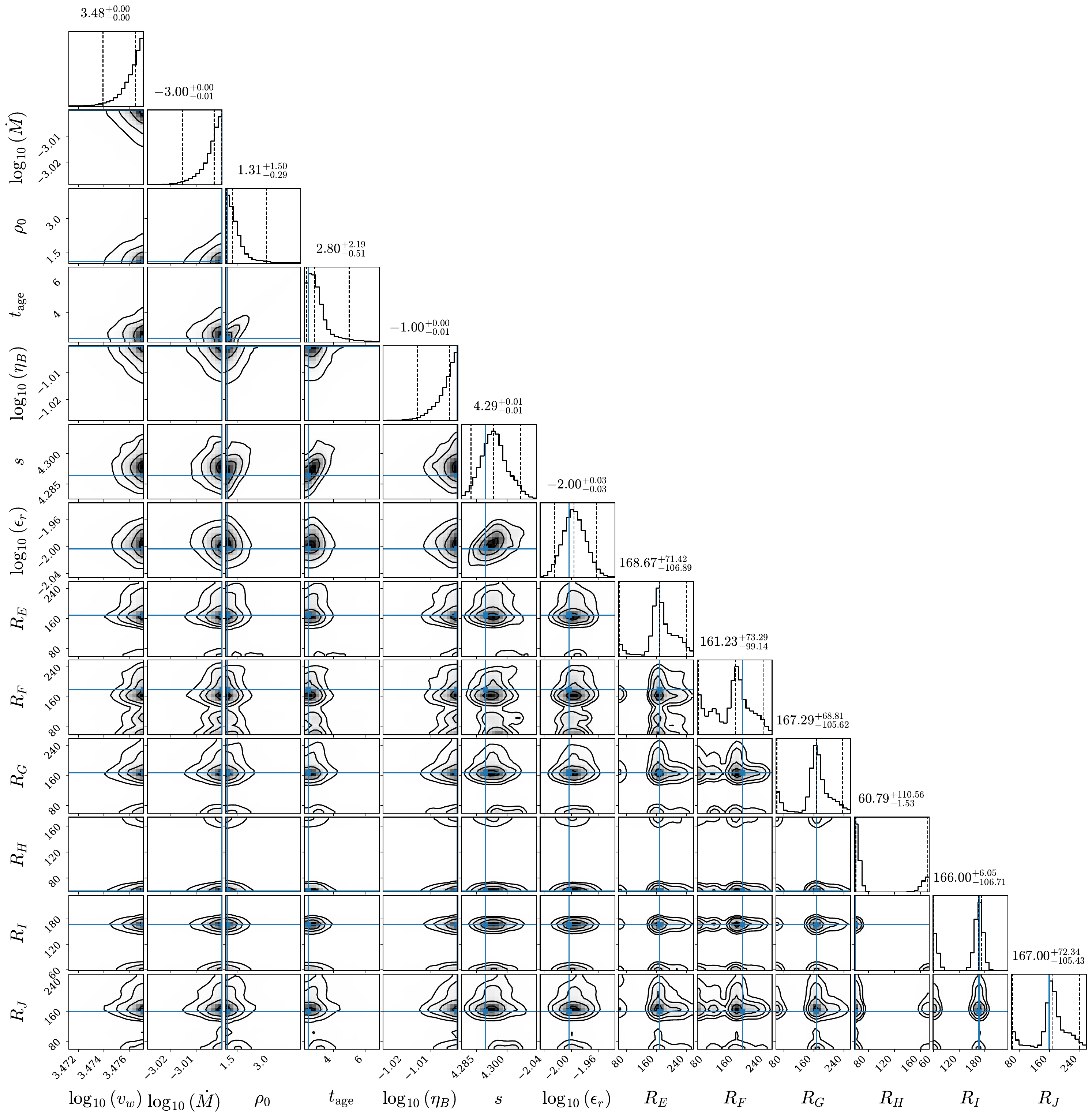}
    \caption{The posterior distribution of the maximised log-likelihood, $\ln \mathscr{L}_i(\mathcal{D}_i \mid \mathcal{P})$, associated with both the spectral and surface brightness profiles of Cyg OB2, derived under the assumption of Kraichnan-type particle diffusion. The blue lines denote the best-fit of the inferred parameter distribution.}
    \label{fig:EKra}
\end{figure}

\begin{figure}
    \centering
    \includegraphics[width=\linewidth]{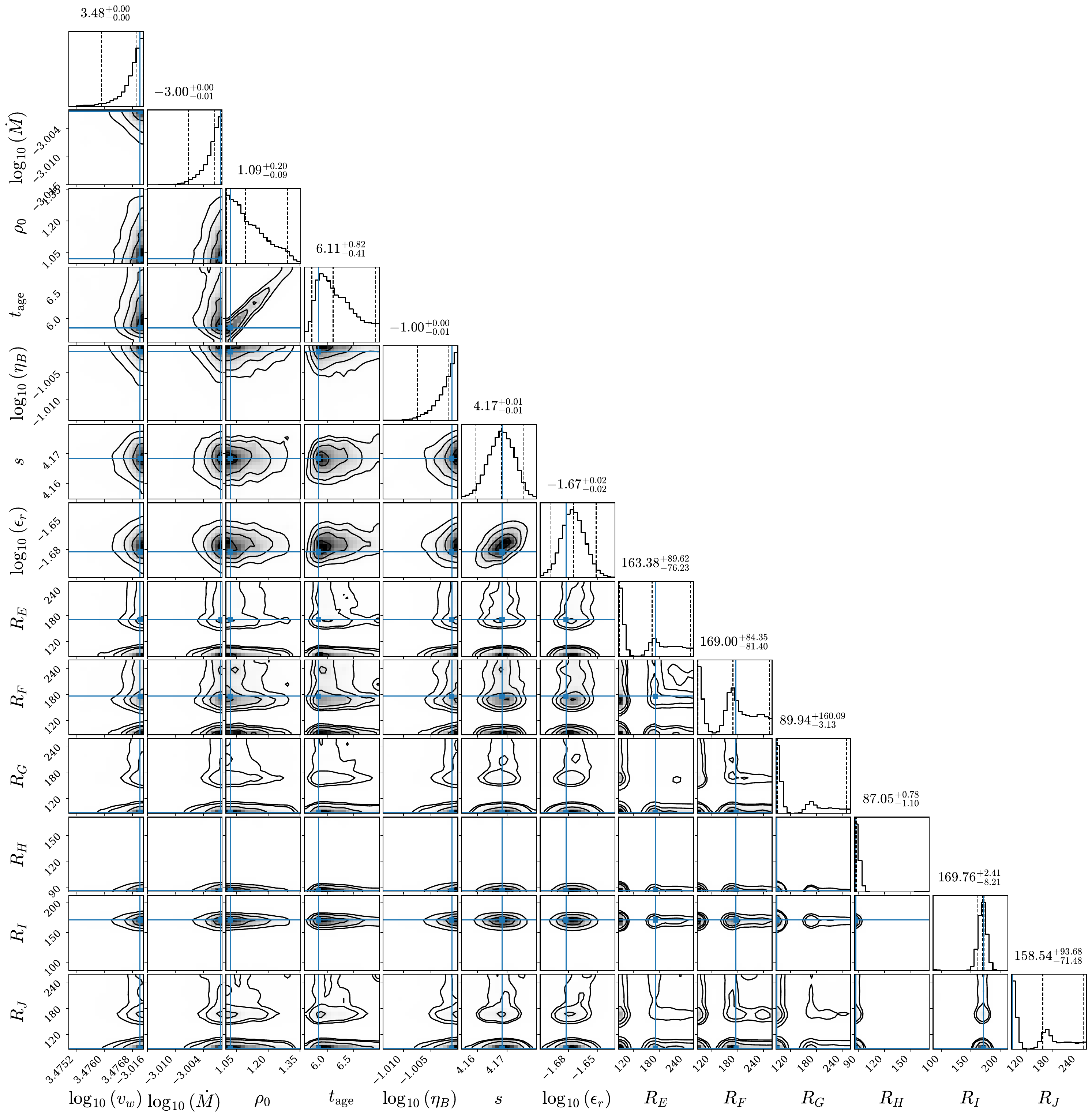}
    \caption{The posterior distribution of the maximised log-likelihood, $\ln \mathscr{L}_i(\mathcal{D}_i \mid \mathcal{P})$, associated with both the spectral and surface brightness profiles of Cyg OB2, derived under the assumption of Kolmogorov-type particle diffusion. The blue lines denote the best-fit of the inferred parameter distribution.}
    \label{fig:EKol}
\end{figure}

To systematically explore the model parameter space, we used \textsc{Nautilus}, a Bayesian nested ensemble sampler designed for complex, multimodal posterior distributions. The method incorporates deep-learning–based components to estimate the Bayesian evidence with high precision. In general, the sampling procedure adopts two convergence criteria, $f_{\rm live}=0.01$ and $n_{\rm eff}=10000$, which respectively specify the target fraction of the evidence remaining in the live point set during exploration and the desired total effective sample size in the subsequent sampling stage. Owing to the pronounced multimodality of the posterior, the sampling efficiency is small $\sim2\times10^{-3}$; therefore, while we retained $f_{\rm live}=0.01$, we relaxed the requirement on the effective sample size to $n_{\rm eff}=1000$. In addition, to mitigate potential bias associated with the exploration phase, we discarded samples drawn during this stage when constructing the final posterior. The resulting posterior distributions are summarised with corner plots showing one-dimensional marginal histograms and two-dimensional credible-region contours, thereby quantifying parameter constraints and correlations across the full posterior space (Figs.~\ref{fig:EBohm}, \ref{fig:EKra}, and \ref{fig:EKol}). In these figures, the dashed lines indicate the $95\%$ credible intervals for each parameter.


\end{document}